%% file: satml_main.tex
\documentclass[conference]{IEEEtran}
\IEEEoverridecommandlockouts

\usepackage{cite}
\usepackage{amsmath,amssymb,amsfonts}
\usepackage{algorithmic}
\usepackage{graphicx}
\usepackage{textcomp}
\usepackage{xcolor}
\usepackage{xcolor}         
\usepackage{graphicx}
\usepackage{subcaption}
\usepackage{amsmath}
\usepackage{amsthm}
\usepackage{bm}
\usepackage{amssymb}
\newtheorem{definition}{Definition}
\newtheorem{theorem}{Theorem}

\usepackage[linesnumbered,ruled,vlined]{algorithm2e}
\usepackage{subfiles} 
\usepackage{tcolorbox}

\usepackage[utf8]{inputenc} 
\usepackage[T1]{fontenc}    
\usepackage{hyperref}       
\usepackage{url}            
\usepackage{booktabs}       
\usepackage{amsfonts}       
\usepackage{nicefrac}       
\usepackage{microtype}      
\usepackage{cleveref}       
\usepackage{lipsum}         
\usepackage{graphicx}
\usepackage[numbers]{natbib}
\usepackage{doi}
\usepackage{enumitem}

\usepackage[linesnumbered,ruled,vlined]{algorithm2e}

\graphicspath{ {./images/} }

\newcommand{\nonrag}{Non-RAG}
\newcommand{\voterag}{VoteRAG}
\newcommand{\dpsparsevoterag}{DPSparseVoteRAG}
\newcommand{\dpvoterag}{DPVoteRAG}

\def\BibTeX{{\rm B\kern-.05em{\sc i\kern-.025em b}\kern-.08em
    T\kern-.1667em\lower.7ex\hbox{E}\kern-.125emX}}
\begin{document}

\title{Privacy-Preserving Retrieval-Augmented Generation with Differential Privacy}

\author{\IEEEauthorblockN{Tatsuki Koga$^*$, Ruihan Wu$^*$\footnote{dsfds}}
\IEEEauthorblockA{University of California, San Diego\\
\textit{$^*$Equal contribution}
}
\and
\IEEEauthorblockN{Zhiyuan Zhang}
\IEEEauthorblockA{University of California, Los Angeles}
\and
\IEEEauthorblockN{Kamalika Chaudhuri}
\IEEEauthorblockA{University of California, San Diego
}
}

\maketitle

\begin{abstract}
\input{abstract}
\end{abstract}

\begin{IEEEkeywords}
component, formatting, style, styling, insert.
\end{IEEEkeywords}

\input{intro}
\input{preliminary}
\input{method}
\input{experiment}
\input{related_work}
\input{conclusion}

\section*{LLM usage considerations}
We primarily use LLMs to refine the grammar and clarity of our writing, while the core ideas and research progress are developed independently through our own study and investigation.

\newpage
\bibliography{ms}
\bibliographystyle{plainnat}
\newpage
\onecolumn
\appendix
\input{appendix}
\end{document}

%% file: abstract.tex
With the recent remarkable advancement of large language models (LLMs), there has been a growing interest in utilizing them in the domains with highly sensitive data that lies outside their training data.
For this purpose, retrieval-augmented generation (RAG) is particularly effective---it assists LLMs by directly providing relevant information from the external knowledge sources.
However, without extra privacy safeguards, RAG outputs risk leaking sensitive information from the external data source.
In this work, we explore RAG under differential privacy (DP), a formal guarantee of data privacy.
The main challenge with differentially private RAG is how to generate long accurate answers within a moderate privacy budget. We address this by proposing an algorithm that smartly spends privacy budget only for the tokens that require the sensitive information and uses the non-private LLM for other tokens. 
Our extensive empirical evaluations reveal that our algorithm outperforms the non-RAG baseline under a reasonable privacy budget of $\epsilon\approx 10$ across different models and datasets.
The code for reproducing our results is at \url{https://github.com/tacchan7412/DPRAG}.

%% file: intro.tex
\section{Introduction}
Large language models (LLMs) have shown a great deal of promise in a variety of applications. 
In particular, a major application of LLMs is in question-answering (QA). The practical adoption of these systems often involves domains whose data is highly sensitive.
For instance, healthcare institutions might want to utilize their internal medical records to provide precise medical information and personal feedback, while legal firms can leverage their case archives to assist clients with legal research and documentation.
One way to achieve such domain-specific QA is through retrieval-augmented generation (RAG)~\citep{chen_reading_2017,lewis_retrieval-augmented_2020, hsia2024ragged}. 
Here, we have a set of domain-specific documents; while answering a question, RAG retrieves a list of relevant documents and inputs them to LLMs as the context. However, even though this is effective for QA, RAG on a sensitive corpus can leak private information about individual documents in the corpus~\cite{zeng_good_2024-2, qi_follow_2024, chaudhari_phantom_2024, peng_data_2024}. 
This is particularly problematic when end users are outside the data-holding entity, e.g., patients interacting with a healthcare institution's RAG system.

Our goal in this paper is to prevent the information leakage of the sensitive external corpus by designing a privacy-preserving RAG system. For this purpose, we use differential privacy (DP)~\cite{dwork_our_2006-1, dwork_calibrating_2006} as a notion of privacy. Differential privacy guarantees privacy by ensuring that the participation of a single person's data does not make much difference to the probability of any output. In our system, we assume that each RAG document comes from a single individual, and our goal is to ensure differential privacy on the eventual answer of the LLM.

There are two aspects of the challenges with designing an effective RAG algorithm under DP.
The first is how to fit differential privacy into the RAG framework, and the second is how to manage the privacy-utility tradeoffs. 
We address the first challenge by proposing an algorithm, \dpvoterag{}, based on the sample-and-aggregate framework in DP~\cite{nissim_smooth_2007}. 
Our algorithm prepares multiple LLM instances, or voters, feeds disjoint partitions of the sensitive corpus to them, and produces output tokens one by one each through the majority vote of the voters' token outputs.
Note, however, that LLMs often output many tokens in response to a question. This is detrimental to privacy---the composition property of differential privacy states that multiple calculations based on the same dataset lead to greater privacy degradation. 
To resolve this challenge, we design another algorithm, \dpsparsevoterag{}, that spends a privacy budget only when we need to.
More specifically, we take advantage of the fact in RAG that LLMs require the sensitive corpus \textit{only} when generating tokens related to the knowledge. When not, outputs from LLMs without any context suffice. 
We formalize this idea with the sparse vector technique in DP~\cite{dwork_complexity_2009, dwork_algorithmic_2014}---when voters agree with the non-private output of the LLM without contexts, we will simply output the non-private one without incurring a privacy budget.
Consequently, our algorithm successfully generates sufficiently long, accurate responses under a reasonable privacy budget.

We conduct extensive experiments with a series of LLMs on multiple benchmarking datasets to evaluate our algorithms.
The results demonstrate that our algorithms are able to enhance the LLMs by RAG while ensuring privacy for the external corpus.
We further show that \dpsparsevoterag{} improves \dpvoterag{} by only spending a privacy budget when necessary and enabling us to generate longer answers within a reasonable privacy budget of $\epsilon\approx 10$. 

%% file: preliminary.tex
\section{Preliminaries \& Problem Setting}
\begin{figure*}[!t]
    \centering
    \includegraphics[width=0.6\textwidth]{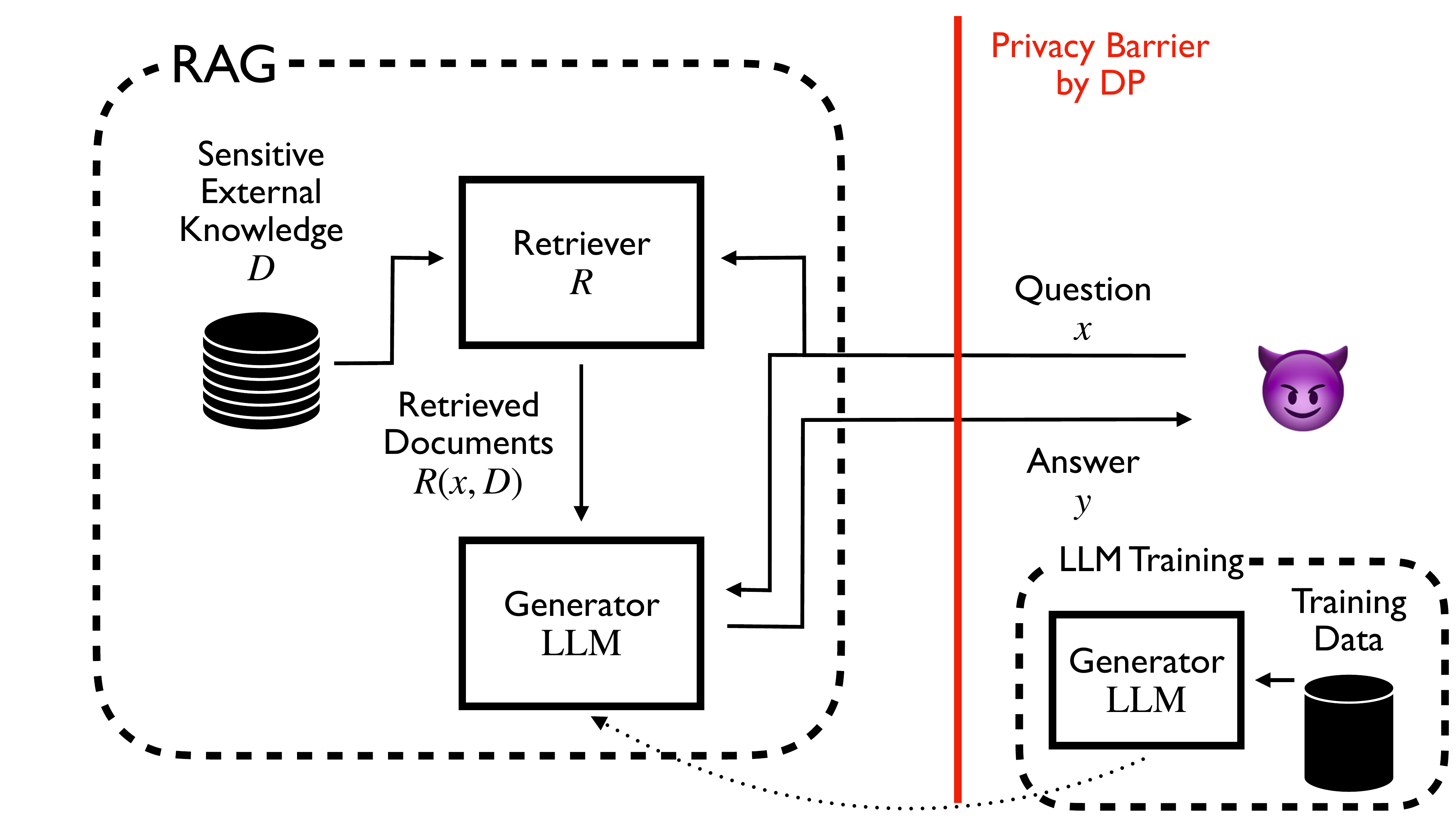}
    \caption{Overview of our problem setting. Note that the LLM in RAG is trained outside the privacy barrier by DP.}
    \label{fig:rag_overview}
\end{figure*}
\subsection{Retrieval-Augmented Generation with Large Language Model}
Retrieval-augmented generation (RAG) is a technique to improve the performance of large language models (LLMs) on knowledge-intensive tasks by providing external knowledge. 
Given a question prompt, a retriever finds relevant documents from the external data source. Then, the relevant documents are added to the prompt as the contexts.
An LLM (or generator) takes the augmented prompt as input and outputs the answer.

More formally, let $x \in \bigcup_{t=1}^{\infty} \mathcal{V}^t$ be a prompt, where $\mathcal{V}$ is some vocabulary. We further let $D$ be a dataset of documents as an external corpus with size $|D|=n$. 
A retriever $R$ finds a subset of $D$, $D_x \subset D$, with size $k$ that is relevant to $x$, i.e., $D_x = R(x, D; k)$.
Finally, an LLM generates an answer $y=\mathrm{LLM}(x, D_x) \in \bigcup_{t=1}^{\infty} \mathcal{V}^t$.
The answer generation can further be decomposed into next-token generation. In particular, for each $t$, the $t$-th token $y_t \in \mathcal{V}$ is generated by $\mathrm{LLM}_t$, which takes $x$, $D_x$, and previously generated tokens $y_{<t}$ as inputs: $y_t=\mathrm{LLM}_t(x,D_x,y_{<t})$.


\subsection{Differential Privacy}
Differential privacy (DP) is a strong cryptographically motivated definition of individual-level privacy. It guarantees that the participation of a single individual in a dataset does not change the probability of any outcome by much. 
In particular, suppose we have two datasets $D$ and $D^\prime$, each consisting of private data from $n$ individuals. We say that $D$ and $D^{\prime}$ are neighboring if they differ in a single individual's private data. 
A randomized algorithm satisfies differential privacy if the output distributions on any pair of neighboring datasets are close.
The formal definition is given as follows.
\begin{definition}[$(\epsilon,\delta)$-Differential Privacy~\cite{dwork_our_2006-1}]
A randomized algorithm $M$ satisfies $(\epsilon, \delta)$-differential privacy if for any two neighboring datasets $D,D^{\prime}$ and for any $S\subseteq \mathrm{range}(M)$,
\begin{align*}
    \Pr [M(D) \in S] \leq \exp (\epsilon) \Pr [M(D^\prime) \in S] + \delta.
\end{align*}
\end{definition}
One of the key properties of DP is composition---sequential runs of differentially private algorithms also satisfy differential privacy. 
The composition property quantitatively captures the intuition that the more we release the information about the sensitive data, the worse the privacy guarantee becomes.
More specifically, suppose $M_1, \ldots, M_T$ are $(\epsilon_0,\delta_0)$-differentially private algorithms, which can be chosen adaptively based on previous outputs. 
Sequential composition theorem~\cite{dwork_our_2006-1} states that the composed sequence of such algorithms guarantee $(T\epsilon_0, T\delta_0)$-differential privacy. 
Furthermore, advanced composition theorem~\cite{dwork_boosting_2010-1, kairouz_composition_2017-1} states that the total privacy guarantee has $\epsilon = \mathcal{O}(\sqrt{T}\epsilon_0)$.

\subsubsection{Sparse Vector Technique}
\label{sec:prelim-svt}
The sparse vector technique~\cite{dwork_complexity_2009, dwork_algorithmic_2014} has originally emerged as the alternative of the composition in DP when we have such a large number of numerical queries that the composition theorem cannot provide a reasonable privacy guarantee but we are only interested in answers above some threshold. 
In such a case, the sparse vector technique algorithm, Sparse, reports whether each (noisy) query answer exceeds the threshold. 
It is shown that the privacy guarantee degrades by the number of queries above the threshold, instead of the total number of queries.
Therefore, we save privacy budget by much when we expect only a few queries will be above the threshold.

We state the AboveThreshold in Algorithm~\ref{alg:at} and state their guarantee as below.

\begin{algorithm}
\caption{AboveThreshold~\citep{dwork_algorithmic_2014}}
\label{alg:at}
\begin{algorithmic}[1]
	\REQUIRE  A private database D, an adaptively chosen
stream of sensitivity 1 queries $f_1$, $\cdots$, and a threshold $\tau$. \ENSURE A stream of responses $a_1$, $\cdots$.
	\STATE Let $\hat{\tau}=\tau + \mathrm{Lap}\left(\frac{2}{\varepsilon}\right)$.
	\FOR{Each query $i$}
	\STATE Let $v_i=\mathrm{Lap}\left(\frac{4}{\varepsilon}\right)$
	\IF{$f_i(D) + v_i\geq \hat{\tau}$}
		\STATE \textbf{Output} $a_i=1$, \textbf{Halt}.
	\ELSE
		\STATE \textbf{Output} $a_i=0$.	
	\ENDIF
	\ENDFOR
\end{algorithmic}
\end{algorithm}

\begin{theorem}
\label{thm:at}
ALgorithm~\ref{alg:at} is $(\varepsilon, 0)$-DP.
\end{theorem}

\subsubsection{Differentially Private Generation via Sample-and-Aggregate}
\label{sec:prelim-dp-vote}
There has been a body of work on generating a token sequence by LLM with DP.
The most common way is to borrow the idea of the sample-and-aggregate framework in DP~\cite{nissim_smooth_2007}. 
To generate a single token, a set of LLMs, each depending on a disjoint subset of the sensitive dataset $D$, generates a token respectively.
The generated tokens form an aggregate histogram of tokens, which is then carefully randomized with noise and only the most frequent token in the noisy histogram is published as the final output.
The repetition of this process along with the composition theorem of DP yields the differentially private token sequence generation. 

\subsection{Problem Setting}
Our goal is to generate an LLM answer to a prompt $x$ with retrieved external knowledge, $D_x = R(x, D; k)$, from a \textit{sensitive} data source $D$ with a differential privacy guarantee.
More specifically, let the sensitive data source $D$ be a collection of individuals' records---one record corresponds to one individual's sensitive data\footnote{It is straightforward to extend the setting to where multiple records correspond to one individual's data by modifying the granularity of neighboring datasets in DP possibly with overhead in privacy-utility tradeoff.}. 
We consider a realistic adversary who does not have direct access to the data source $D$ but has a capability of querying to the RAG system with any prompt $x$.
We further assume that the LLM used in the RAG system is a copy of publicly available LLMs and is already pre-trained (and fine-tuned) with data \textit{disjoint} from the sensitive data source $D$. That is, having access to the LLM parameters and/or pre-training (and fine-tuning) data does not provide any information on the sensitive data source $D$.
To this end, we aim to formally guarantee that given any question $x$, a randomized LLM generation algorithm with RAG, $\mathrm{LLM}_\mathrm{priv}(x, R(x,D; k))$ satisfies $(\epsilon, \delta)$-differential privacy w.r.t the external knowledge data source $D$.
We present the figure for this problem setting in Figure~\ref{fig:rag_overview}.

%% file: method.tex
\section{Differentially Private Retrieval-Augmented Generation with Sparse Vector Technique}
\begin{figure*}[!t]
    \centering
    \includegraphics[width=\textwidth]{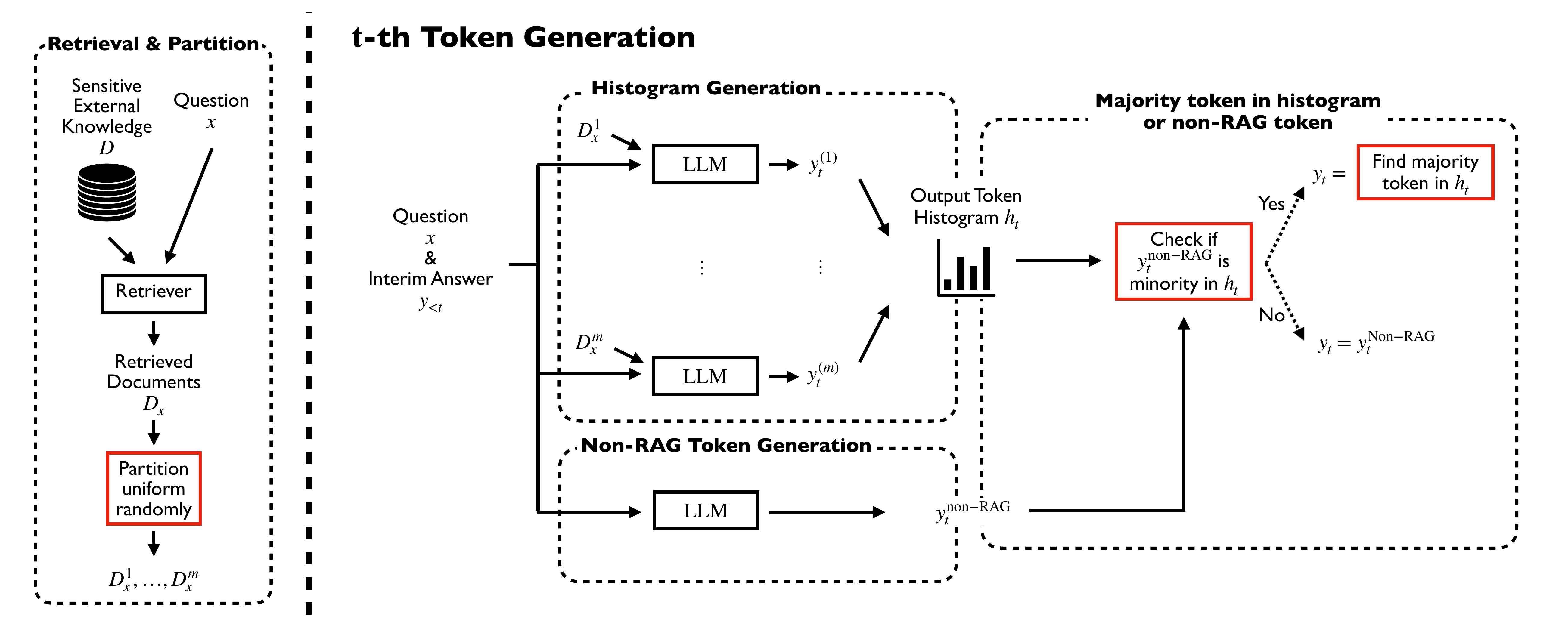}
    \caption{
    Overview of \dpsparsevoterag{}. 
    It first retrieves the relevant documents to a question $x$ from sensitive external knowledge $D$ and partitions them uniform randomly (left).
    When generating $t$-th token (right), it takes a question $x$ and an interim answer up to $t-1$-th tokens, $y_{<t}$, as inputs and outputs a new token $y_t$.
    Components enclosed in red indicate that operations involve randomness for the privacy guarantee.
    \dpvoterag{} works the same until computing the output token histogram $h_t$, but it immediately finds the majority token in $h_t$ afterwards.
    }
    \label{fig:dprag_overview}
\end{figure*}
Our differentially private RAG algorithm consists of two main components---DP voting for the single-token generation and efficient privacy budget spending by leveraging
the sparse vector technique combined with the utilization of LLMs without any relevant documents provided.
These two components enable us to generate answers that incorporate external knowledge while guaranteeing a reasonable level of differential privacy.
We start from our algorithm with the first component alone, and then extend it to include the second component. 
The graphical overview of our algorithm is presented in Figure~\ref{fig:dprag_overview}.

\subsection{DPVoteRAG: Differentially Private Voting Algorithm for RAG}



\begin{algorithm*}[tb]
\caption{DPVoteRAG}
\label{alg:dp-vote}
\begin{algorithmic}[1]
\REQUIRE Prompt $x$, External data source $D$, Generator $\mathrm{LLM}$, Retriever $R$, 
         \# of voters $m$, \# of retrieval per voter $k$,
         Per-token privacy budget $(\epsilon_\mathrm{token}, \delta_\mathrm{token})$,
         Total privacy budget $(\epsilon_\mathrm{total}, \delta_\mathrm{total})$
\ENSURE Private answer $y$
\STATE $T_{\max} \gets$ maximum \# of tokens to generate based on $(\epsilon_\mathrm{token}, \delta_\mathrm{token})$ and $(\epsilon_\mathrm{total}, \delta_\mathrm{total})$
\STATE \texttt{\small \COMMENT{Retrieval and random partition of the relevant documents}}
\STATE $D_x\gets$  Retrieve $m k$ most relevant documents $R(x, D; mk)$.
\STATE $D_x^{1}, \cdots, D_x^{m}\gets$ Uniformly randomly partition $D_x$ into $m$ disjoint subsets.
\FOR{$t \leftarrow 1$ \textbf{to} $T_{\max}$}
    \STATE \texttt{\small \COMMENT{Generating the token histogram with the sensitive documents}} 
    \FOR{$i \leftarrow 1$ \textbf{to} $m$}
        \STATE $y_t^{(i)} \gets \mathrm{LLM}_t(x,D_x^{i},y_{<t})$ 
    \ENDFOR
    \STATE $h_t \gets$ Build a histogram of tokens from $y_t^{(1)}, \ldots, y_t^{(m)}$
    \STATE  \texttt{\small \COMMENT{Producing a token from the histogram privately}}
    \STATE $y_t \gets \mathrm{LimitedDomain}(h_t, \epsilon_\mathrm{token}, \delta_\mathrm{token})$ 
    \STATE \texttt{\small \COMMENT{Halting if end of sequence}}
    \IF{$y_t = \texttt{<EOS>}$}
        \STATE \textbf{return} $(y_1,\ldots,y_t)$ 
    \ENDIF
\ENDFOR
\STATE \textbf{return} $(y_1,\ldots,y_{T_{\max}})$

\end{algorithmic}
\end{algorithm*}

By the nature of retrieval in RAG---retrieving relevant documents for a question, the LLM outputs can depend on a sensitive individual's document.
Therefore, our algorithmic design needs to relax the dependency of a single individual's document on the output, while exploiting the external data source, to achieve a reasonable privacy-utility tradeoff.
Inspired by the differentially private generation via sample-and-aggregate framework, we present a differentially private voting algorithm for RAG---\textbf{DPVoteRAG}.

Given a prompt $x$ and external data source $D$, \dpvoterag{} first retrieves $mk$ documents as $D_x$. Then it makes uniformly randomly partitions $D_x$ into $m$ disjoint datasets $D_x^{1},\ldots,D_x^{m}$ and each subset has exact size $k$.
Then, for each $i=1,\ldots, m$, it feeds $k$ documents $D_x^{i}$ into the LLM along with the original prompt $x$, and generates a next token.
It collects these tokens to form a histogram over the vocabulary.
It remains to privately choose the most frequent element from the histogram.
While it is generally hard to do so when the histogram dimension is large as in our setting, e.g., the vocabulary size of OPT~\cite{zhang_opt_2022-1} is 50272, there is a line of work in the community to overcome this difficulty.
Following the work by~\citet{hong2024dpopt}, we integrate the LimitedDomain mechanism~\cite{durfee_practical_2019-1} into our algorithm. The mechanism enables us to reduce the histogram dimension significantly with some cost in a privacy budget and thus achieve a better privacy-utility tradeoff.
By its design, the LimitedDomain mechanism possibly outputs the designed null token. 
In such a case, we halt the algorithm, or equivalently, regard that it outputs the end of sequence token.
\footnote{We find that by choosing the appropriate size of reduced dimension, the LimitedDomain mechanism in our experiment rarely outputs the null token.}
Finally, we append the chosen token to the next input to the LLM.
We repeat this process until we see the end of sequence token chosen or reach the maximum number of token generation, which is computed in advance from the per-token and total privacy budget
\footnote{The maximum number of token generation is computed as follows. 
We first calculate the maximum numbers of composition with the sequential and advanced composition theorem~\cite{dwork_algorithmic_2014} under the per-token privacy budget $(\epsilon_\mathrm{token},\delta_\mathrm{token})$ and total privacy budget $(\epsilon_\mathrm{total},\delta_\mathrm{total})$.
Then, we take the maximum of two numbers of possible composition.}.
We present the concrete algorithm in Algorithm~\ref{alg:dp-vote}.
The formal privacy analysis is as follows.
\begin{theorem}
\label{thm:dp}
For any question $x$, \dpvoterag{} satisfies $(\epsilon_\mathrm{total}, \delta_\mathrm{total})$-DP w.r.t. the external data source $D$.
\end{theorem}
The guarantee simply follows from the property of uniformly random partition, the privacy guarantee of the LimitedDomain mechanism and the composition theorem used to compute $T_{\max}$. 

\begin{proof}
Let's first consider steps 3 and 4 in Algorithm~\ref{alg:dp-vote}. Suppose $L_x$ is the list of documents in $D_x$ ranked by the relevance. One way to uniformly randomly split $D_x$ into $m$ disjoint subsets $D_x^{1}, \cdots, D_x^{m}$ is that: given a ranked list of documents $L_x=(d_1, \cdots, d_{mk})$, we randomly permute this list by $\pi$ to $L_x^{\pi}=(d_{\pi(1)}, \cdots, d_{\pi(mk)})$ and let $D_x^{i}:=\{d_{\pi((i-1)k+1)}, \cdots, d_{\pi(ik)}\}$. The process from $D_x$ to $L_x$ is deterministic, and the remaining of the algorithm is independent of $D$ given $L_x^{\pi}$. Therefore, we can equivalently denote the outcome of Algorithm~\ref{alg:dp-vote} as $\mathcal{A}(L_x^{\pi})$.

For any two neighboring datasets $D$ and $D'$, the retrieved datasets are $D_x=R(x, D; mk)$ and $D_x'=R(x, D'; mk)$ and we denote $L_x=(d_1, \cdots, d_{mk})$ and $L_x'=(d_1', \cdots, d_{mk}')$. We only need to show for any set of outcomes $S$, $\Pr_{A, \pi} [A(L_x^{\pi}) \in S] \leq \exp (\epsilon) \Pr_{A, \pi} [A((L_x')^{\pi}) \in S] + \delta.$ First of all, $D_x$ and $D_x'$ have at most one different document (without considering the order). Therefore we can define another list of documents $L''_x=(d_1'', \cdots, d_{mk}'')$, such that $L''_x$ is some ranking of $D_x’$ and it differs at most one position from $L$. Notice that $L'_x$ and $L''_x$ are only different at orders. Therefore $(L_x')^{\pi}$ and $(L_x'')^{\pi}$ have same distributions and as a consequence $A((L_x')^{\pi})$ and $A((L_x'')^{\pi})$ have same distributions. 

Thus, the remaining is to prove for any set of outcomes $S$, $\Pr_{A, \pi} [A(L_x^{\pi}) \in S] \leq \exp (\epsilon) \Pr_{A, \pi} [A((L_x'')^{\pi}) \in S] + \delta.$ We can actually prove a stronger conclusion $\Pr_{A} [A(L_x^{\pi}) \in S] \leq \exp (\epsilon) \Pr_{A} [A((L_x'')^{\pi}) \in S] + \delta.$ It is because $L_x^{\pi}$ and $(L_x'')^{\pi}$ differ at most one position and therefore at most one subset in step 4 is different given  $L_x^{\pi}$ or $(L_x'')^{\pi}$. 
This means that the histogram in step 10 differs at most one token.
The guarantees of LimitedDomain and the composition theorem of DP together imply $\Pr_{A} [A(L_x^{\pi}) \in S] \leq \exp (\epsilon) \Pr_{A} [A((L_x'')^{\pi}) \in S] + \delta.$
\end{proof}

\subsection{DPSparseVoteRAG: Differentially Private Voting Algorithm for RAG with Sparse Vector Technique}

\begin{algorithm*}[!t]
\caption{DPSparseVoteRAG}
\label{alg:dp-sparse-vote}
\begin{algorithmic}[1]
\REQUIRE Prompt $x$, External data source $D$, Generator $\mathrm{LLM}$, Retriever $R$, 
         \# of voters $m$, \# of retrieval per voter $k$,
         Per-token privacy budget $(\epsilon_\mathrm{token}, \delta_\mathrm{token})$,
         Total privacy budget $(\epsilon_\mathrm{total}, \delta_\mathrm{total})$,
         Threshold $\tau$, Maximum \# of output tokens (regardless of privacy) $T_{\max}$ 
\ENSURE Private answer $y$
\STATE  \texttt{\small \COMMENT{Privacy budget setup}}
\STATE $(\epsilon_\mathrm{token-RAG}, \delta_\mathrm{token-RAG}) \gets (\epsilon_\mathrm{token}/2, \delta_\mathrm{token})$, $\epsilon_\mathrm{token-Lap} \gets \epsilon_\mathrm{token}/2$
\STATE $c_{\max} \gets$ maximum \# of tokens to generate \textit{privately} based on $(\epsilon_\mathrm{token}, \delta_\mathrm{token})$ and $(\epsilon_\mathrm{total}, \delta_\mathrm{total})$
\STATE $c \gets c_{\max}$, $\hat{\tau} \gets \tau + \mathrm{Lap}(\nicefrac{2}{\epsilon_\mathrm{token-Lap}})$
\STATE
 \texttt{\small \COMMENT{Retrieval and random partition of the relevant documents}}
\STATE $D_x\gets$  Retrieve $m k$ most relevant documents $R(x, D; mk)$.
\STATE $D_x^{1}, \cdots, D_x^{m}\gets$ Uniformly randomly partition $D_x$ into $m$ disjoint subsets.
\FOR{$t \gets 1$ \textbf{to} $T_{\max}$}
    \STATE \texttt{\small \COMMENT{Generating the non-private token and token histogram with the sensitive documents}} 
    \STATE $y_t^\mathrm{non-RAG} \gets \mathrm{LLM}_t(x,`` '',y_{<t})$
    \FOR{$i \leftarrow 1$ \textbf{to} $m$}
        \STATE $y_t^{(i)} \gets \mathrm{LLM}_t(x,D_x^{i},y_{<t})$ 
    \ENDFOR
    \STATE $h_t \gets$ Build a histogram of tokens from $y_t^{(1)}, \ldots, y_t^{(m)}$
    \STATE  \texttt{\small \COMMENT{Producing a token from the histogram privately only when $y_t^\mathrm{non-RAG}$ is uncommon in $h_t$}}
    \STATE $a_t \gets$ Extract a count of $h_t$ at $y_t^\mathrm{non-RAG}$
    \IF{$a_t + \mathrm{Lap}(\nicefrac{4}{\epsilon_\mathrm{token-Lap}}) \leq \hat{\tau}$}
        \STATE $y_t \gets \mathrm{LimitedDomain}(h_t, \epsilon_\mathrm{token-RAG}, \delta_\mathrm{token-RAG})$
        \STATE \texttt{\small \COMMENT{The privacy budget is only consumed when $y_t$ is from the histogram}}
        \STATE $c \gets c-1$, $\hat{\tau} \gets \tau + \mathrm{Lap}(\nicefrac{2}{\epsilon_\mathrm{token-Lap}})$
    \ELSE
        \STATE $y_t \gets y_t^\mathrm{non-RAG}$
    \ENDIF
    \STATE \texttt{\small \COMMENT{Halting if end of sequence or the privacy budget has been exhausted}}
    \IF{$y_t = \texttt{<EOS>}$ or $c = 0$}
        \STATE \textbf{return} $(y_1,\ldots,y_t)$
    \ENDIF
\ENDFOR
\STATE \textbf{return} $(y_1,\ldots,y_{T_{\max}})$

\end{algorithmic}
\end{algorithm*}

The main drawback of the aforementioned algorithm is that we need to spend a non-negligible amount of privacy budget for each token to guarantee its quality.
This prevents our algorithm from generating longer answers---sometimes it can halt before it generates the actual answers due to privacy budget shortage.
More concretely, consider the following question-answering example:
\begin{tcolorbox}[
    colback=gray!5,
    colframe=blue!75!black,
    fonttitle=\bfseries
]
\textbf{Question}: what type of literature is the great gatsby

\textbf{Ground Truth Answer}: novel
\end{tcolorbox}
Here are possible outputs from (non-private) RAG and our \dpvoterag{} given the retrieved documents.
\begin{tcolorbox}[
    colback=gray!5,
    colframe=blue!75!black,
    fonttitle=\bfseries
]
\textbf{RAG Output:} The Great Gatsby is a novel written by American author F. Scott Fitzgerald.

\textbf{DPVoteRAG Output:} The Great Gatsby is a
\end{tcolorbox}
While non-private RAG correctly answers the question, due to the pre-fixed total privacy budget, DPVoteRAG can only output 5 words and thus it fails to output the ground truth answer, \textit{novel}.

However, having a closer look at our voting algorithm, we observe that there is room for improvement.
When generating the 3rd word, \textit{Gatsby}, every input of the LLM contains \textit{The Great}, 1st and 2nd previously output words, and \textit{the great gatsby}, a part of the question, even though the provided retrieved documents are different.
Thus, the LLM should successfully generate \textit{Gatsby} without access to the sensitive information. Ideally, we should not spend a privacy budget for such a word.

We address this by incorporating the sparse vector technique into our voting algorithm, yielding our improved algorithm \textbf{DPSparseVoteRAG}.
In particular, before we apply private voting among generated tokens, we check if the generated tokens coincide with the token generated by the LLM without retrieved documents appended, i.e., the input is composed of the prompt and previously generated tokens only.
We continue to the voting only when they do not coincide. Otherwise, we use the LLM output without retrieved documents.
It is shown from the analysis of the sparse vector technique that the consumed privacy budget scales with the number of times that it uses the private voting, not with the total number of generated tokens.
Consequently, the resulting algorithm, shown in Algorithm~\ref{alg:dp-sparse-vote}, enables us to spend a privacy budget only when it needs sensitive information.
We note that as a result of this change in our algorithm, we compute the maximum number of tokens to generate \textit{with private voting}, $c_{\max}$, from the per-token privacy budget $(\epsilon_\mathrm{token},\delta_\mathrm{token})$ and total privacy budget $(\epsilon_\mathrm{total},\delta_\mathrm{total})$, instead of the maximum number of token generation, $T_{\max}$, as in \dpvoterag{} but in the same way to compute $T_{\max}$.
The formal privacy analysis of this algorithm is as follows.
The guarantee holds due to the privacy guarantee of the LimitedDomain mechanism and the AboveThreshold algorithm~\cite{dwork_algorithmic_2014}.
\begin{theorem}
For any question $x$, \dpsparsevoterag{} satisfies $(\epsilon_\mathrm{total}, \delta_\mathrm{total})$-DP w.r.t. the external data source $D$.
\end{theorem}
\begin{proof}
Similar to the proof of Theorem~\ref{thm:dp}, we only need to prove that if $\mathcal{D}_x = (D_x^1, \cdots, D_x^m)$ at step 7 and $\mathcal{D}_x' = (D_x^{1'}, \cdots, D_x^{m'})$ differ at most one set $D_x^i$, $\Pr(y_1, \cdots, y_{T_{\max}})\leq \varepsilon\cdot \Pr(y_1', \cdots, y_{T_{\max}}')+\delta$.

Denote $t_{\gamma}$ as the first time step that holds $c=\gamma$ at the beginning of this time step, e.g. $t_{c_{\max}}=1$.
We can first prove that 
$\Pr(y_{t_{\gamma}\leq t< t_{\gamma-1}}|\mathcal{D}_x, y_{t<t_{\gamma}})\leq \varepsilon_{\rm token}\cdot \Pr(y_{t_{\gamma}'\leq t< t_{\gamma-1}'}'|\mathcal{D}_x', y_{t<t_{\gamma}'}')+\delta_{\rm token}$
if (1) $t_{\gamma}=t_{\gamma}'$ and (2) $y_{t<t_{\gamma}}=y_{t<t_{\gamma}}'$.

This can be proved by two parts. First, because $t_{\gamma}=t_{\gamma}'$,
$$\Pr(t_{\gamma-1}|\mathcal{D}_x, y_{t<t_{\gamma}})\leq \varepsilon_{\rm token}/2\cdot \Pr(t_{\gamma-1}|\mathcal{D}_x', y_{t<t_{\gamma}'}').$$
This is implied by applying the sparse vector technique presented in Algorithm~\ref{alg:at} and Theorem~\ref{thm:at} to analyze our algorithm (step 17-20).
Furthermore, if $t_{\gamma-1}=t_{\gamma-1}'$, 
$$\Pr(y_{t_{\gamma}\leq t< t_{\gamma-1}}|\mathcal{D}_x, y_{t<t_{\gamma}}) = \Pr(y_{t_{\gamma}\leq t < t_{\gamma-1}}^{non-RAG}, y_{t_{\gamma-1}}^{DP}|\mathcal{D}_x, y_{t<t_{\gamma}})$$
$$ = \Pr(y_{t_{\gamma}'\leq t < t_{\gamma-1}'}^{non-RAG}, y_{t_{\gamma-1}'}^{DP}|\mathcal{D}_x, y_{t<t_{\gamma}'}') $$
$$\leq \varepsilon_{\rm token}/2\cdot \Pr(y_{t_{\gamma}'\leq t < t_{\gamma-1}'}^{non-RAG}, y_{t_{\gamma-1}}^{DP'}|\mathcal{D}_x', y_{t<t_{\gamma}'}')+\delta_{\rm token}/2$$
$$
=\Pr(y_{t_{\gamma}'\leq t< t_{\gamma-1}'}'|\mathcal{D}_x', y_{t<t_{\gamma}}') 
$$
where the first and the last equality come from the definition of the algorithm (step 17-23), the second equality holds because we assume $t_{\gamma}=t_{\gamma}'$, $y_{t<t_{\gamma}}=y_{t<t_{\gamma}}'$ and $t_{\gamma-1}=t_{\gamma-1}'$, and the inequality comes from the DP guarantee by the LimitedDomain mechanism.

Lastly, our algorithm must stop before $c=0$, means that our algorithm is a composition of at most $c_{\max}$ steps of $(\varepsilon_{\rm token}, \delta_{\rm token})$-DP.
As shown in step 3 in our algorithm, $c_{\max}$ is picked to guarantee that the composition of $c_{\max}$ steps of $(\varepsilon_{\rm token}, \delta_{\rm token})$-DP is $(\varepsilon_{\rm total}, \delta_{\rm total})$-DP.
Therefore, our algorithm is $(\varepsilon_{\rm total}, \delta_{\rm total})$-DP.
\end{proof}

%% file: experiment.tex
\section{Experiment}
We investigate how our differentially private voting RAG algorithms (Algorithms~\ref{alg:dp-vote} and~\ref{alg:dp-sparse-vote}) work. Specifically, we ask the following questions:
\begin{enumerate}[nosep]
    \item How do our algorithms improve the accuracy of question-answering over non-RAG LLM while ensuring a formal privacy guarantee?
    \item Is DPSparseVoteRAG (Algorithm~\ref{alg:dp-sparse-vote}) always a better choice than DPVoteRAG (Algorithm~\ref{alg:dp-vote})?
    \item Is there any useful guidance of choosing hyperparameters $m$ (the number of voters) and $\epsilon_\mathrm{token}$?
    \item How do our algorithm protect against empirical privacy attack?
\end{enumerate}
We study each question through extensive evaluations on the well-used benchmarking datasets with multiple LLMs.

\subsection{Methodology}
\paragraph{Datasets.}
We use two question-answering benchmarking datasets for RAG: \textbf{Trivia}~\cite{joshi_triviaqa_2017} and \textbf{Natural Question (NQ)}~\cite{kwiatkowski_natural_2019}. 
Each dataset consists of a list of pairs of question and answer lists, i.e., every question can have multiple answers.
By following the standard evaluations in RAG~\cite{chen_reading_2017,lewis_retrieval-augmented_2020, hsia2024ragged}, we use the Wikipedia dataset as the external data source from which a retriever finds relevant documents.
For each dataset, we use a subset of 100 questions to manage the computational overhead
\footnote{We first filter out questions to less than 20 ground truth documents in the Wikipedia dataset. If a question relates with only a few documents, DP algorithms will likely fail since replacing a document would change the output a lot. Then, we split the remaining questions into 4 bins with 20--29, 30--39, 40--49 and 50--59 relevant documents and sample 25 questions from each bin.}. 

In addition, we experiment with a realistic privacy-sensitive application, where the external corpus contains inherently private information.
Chatdoctor Questions~\citep{li2023chatdoctor} consist of QA interactions between patients and doctors in the healthcare domain. We sample 100 patient questions from the original dataset as our test set. 
The external dataset consists of the remaining QA pairs from the original ChatDoctor dataset, excluding the 100 patient questions used for testing.
Here is a document example:
\begin{tcolorbox}[
    colback=gray!5,
    colframe=blue!75!black,
    fonttitle=\bfseries
]
\textit{Patient's description}: My son fell from bed heads on, and didn t vomit or pass out. However, we put him to sleep as this was his sleep time. After two hours he woke up, and we felt he had a fever. We gave him brufen...
\#\#\# \textit{Doctor's answer}: Hi, according to me, I think since the kid fell and did not have vomiting, indicates there is no concussion. The fever is incidental, which can occur after a fall. This is normal. Give the kid paracetamol, probably every 6 hours once. If even after two days fever does not subside, kindly visit your pediatrician.... Take care.
\end{tcolorbox}

\paragraph{Models.}
The retriever we use is the Dense Passage Retriever (DPR)~\cite{karpukhin_dense_2020} which is built on top of BERT~\cite{devlin_bert_2019-1}. It finds relevant documents that are close to the question in the embedding space produced by BERT.
We compare the following generator LLMs: OPT (1.3B)~\cite{zhang_opt_2022-1}, Llama 3.1 (8B)~\cite{dubey2024llama}, and Pythia (1.4B)~\cite{biderman_pythia_2023-1}.
We additionally report the result of OPT (2.7B), Llama 3.2 (1B), and GPT2-XL~\cite{radford_language_nodate} in Appendix~\ref{appendix:additional-exp}.

\paragraph{Algorithms.}
We compare our algorithms, \textbf{DPVoteRAG} (Algorithm~\ref{alg:dp-vote}) and \textbf{DPSparseVoteRAG} (Algorithm~\ref{alg:dp-sparse-vote}), with two baseline algorithms. 
One baseline algorithm is \textbf{Non-RAG} where we only provide a question to the LLM without any relevant documents appended as a prompt. In order for our algorithms to be useful, they have to outperform this baseline.
The other is \textbf{VoteRAG} where we carry out the same voting procedure as our algorithms but choose the most frequent token across voters non-privately---the most frequent token is always chosen as the next token to generate. For each number of voters, the result of this baseline serves as the upper bound of our DP algorithms.

\paragraph{Experimental Setup.}
We observe the results under multiple total privacy budgets, $(\epsilon_\mathrm{total},\delta_\mathrm{total})$. More specifically, we sweep $\epsilon_\mathrm{total} = 2$ to $40$ and set $\delta_\mathrm{total}=10^{-4}$.
Furthermore, we consider different per-token privacy budgets for our private algorithms: $\epsilon_\mathrm{token} = 1,2,5$ and $\delta_\mathrm{token}=10^{-5}$.
We consider the number of voters $m$ of $10$, $20$, $30$, $40$, and $50$ for VoteRAG, and $30$, $40$, and $50$ for DPVoteRAG and DPSparseVoteRAG so as to ensure reasonable privacy-utility tradeoff and computational overhead. 
For DPSparseVoteRAG, we set the threshold $\tau$ to be half of the number of voters, i.e., $\tau=m/2$.
When we use the LimitedDomain mechanism to privately choose the most frequent token, we set their parameter $\bar{k}$ to be the number of voters, where $\bar{k}$ is the limited size of the domain to which we add the Gumbel noise.
For voting algorithms, each voter receives 1 relevant document, i.e., $k=1$.
The utility evaluation metric is the match accuracy~\cite{mallen_when_2023,asai_self-rag_2023-1,schick_toolformer_2023,zhang_retrievalqa_2024-1} which measures if the prediction to 
a question contains any of its answers.

\subsection{Main Results}
\begin{figure*}[t]
    \centering
    \includegraphics[width=0.3\textwidth]{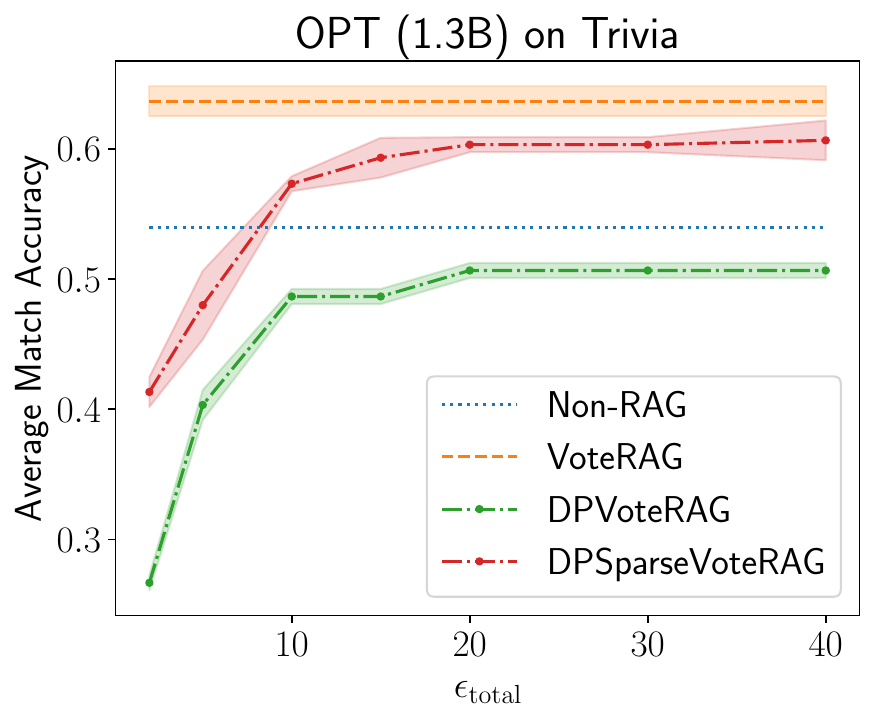}
    \hfill
    \includegraphics[width=0.3\textwidth]{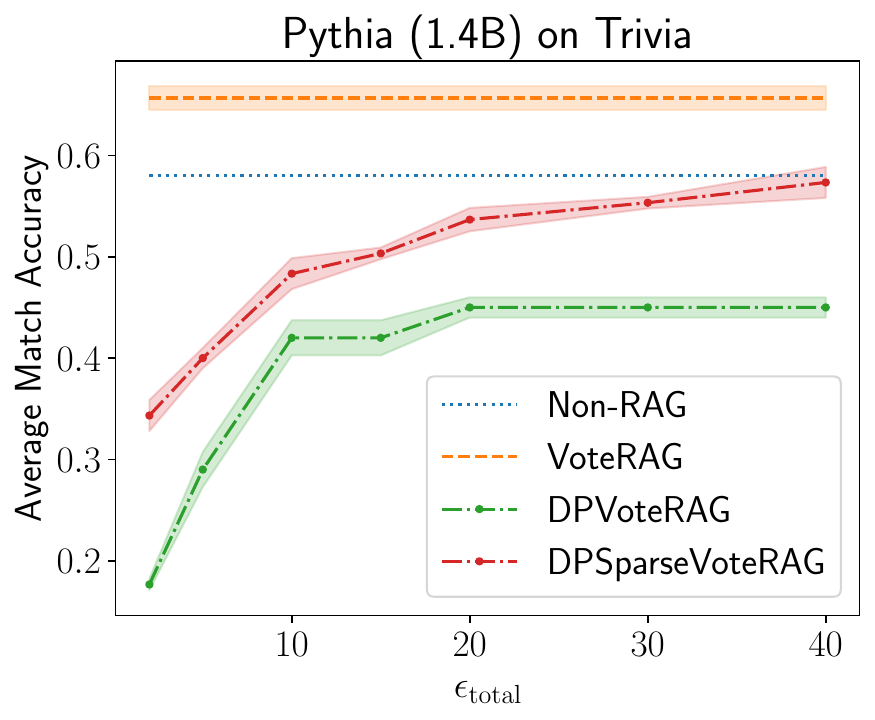}
    \hfill
    \includegraphics[width=0.3\textwidth]{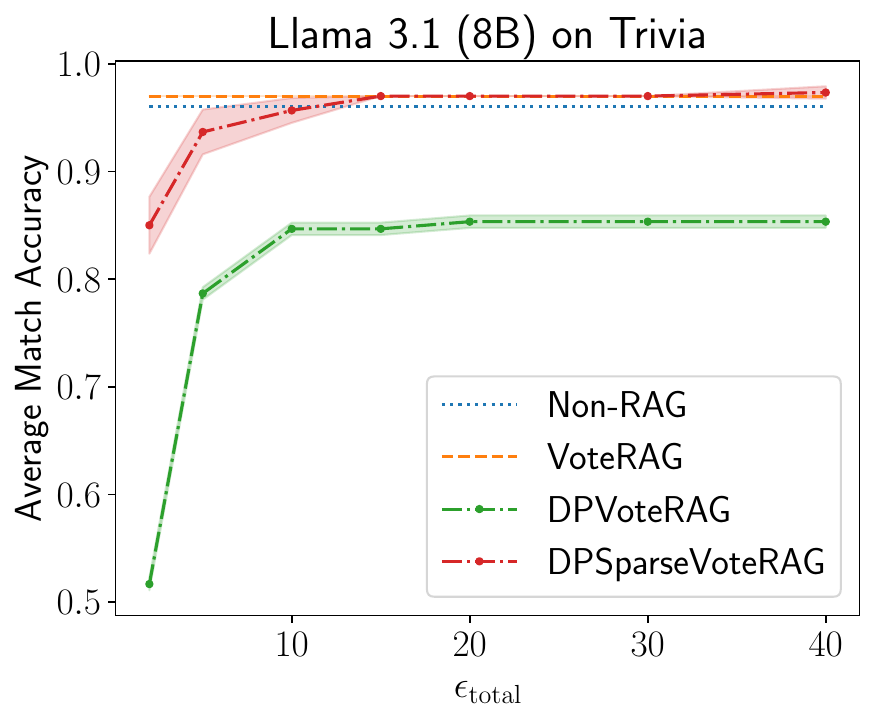}
    
    
    \includegraphics[width=0.3\textwidth]{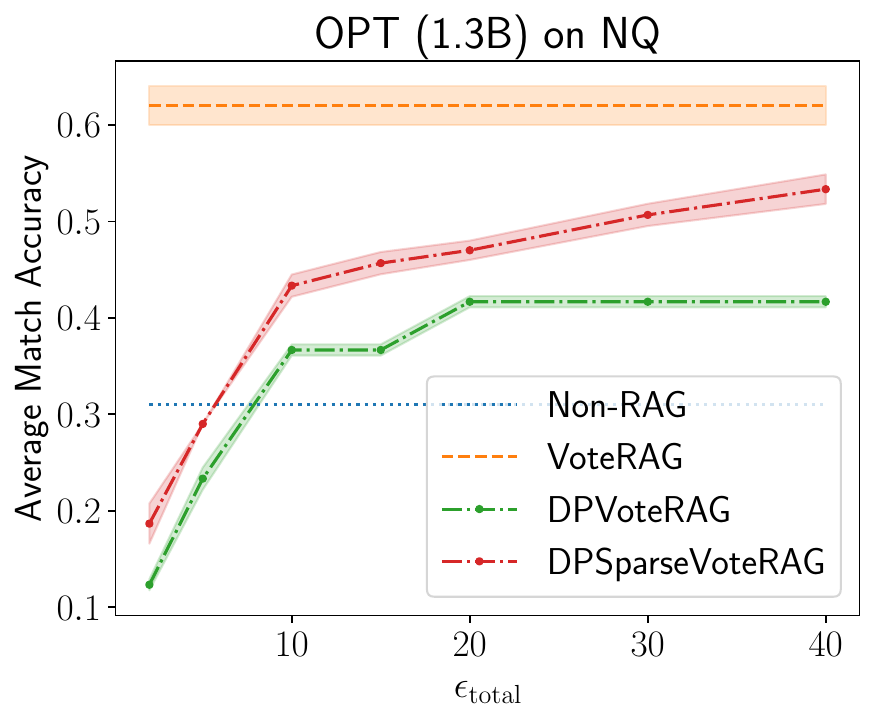}
    \hfill
    \includegraphics[width=0.3\textwidth]{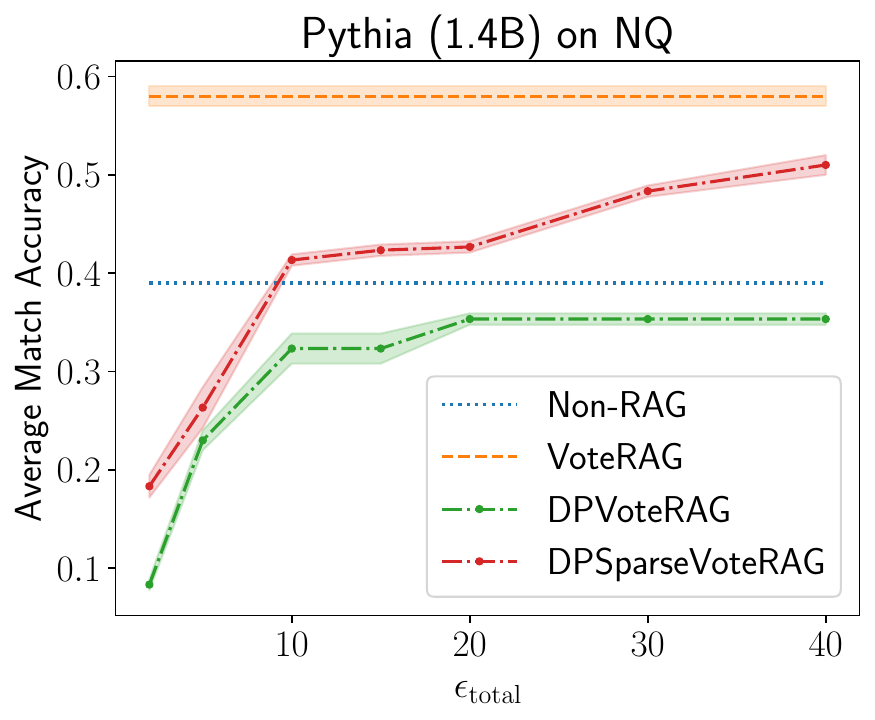}
    \hfill
    \includegraphics[width=0.3\textwidth]{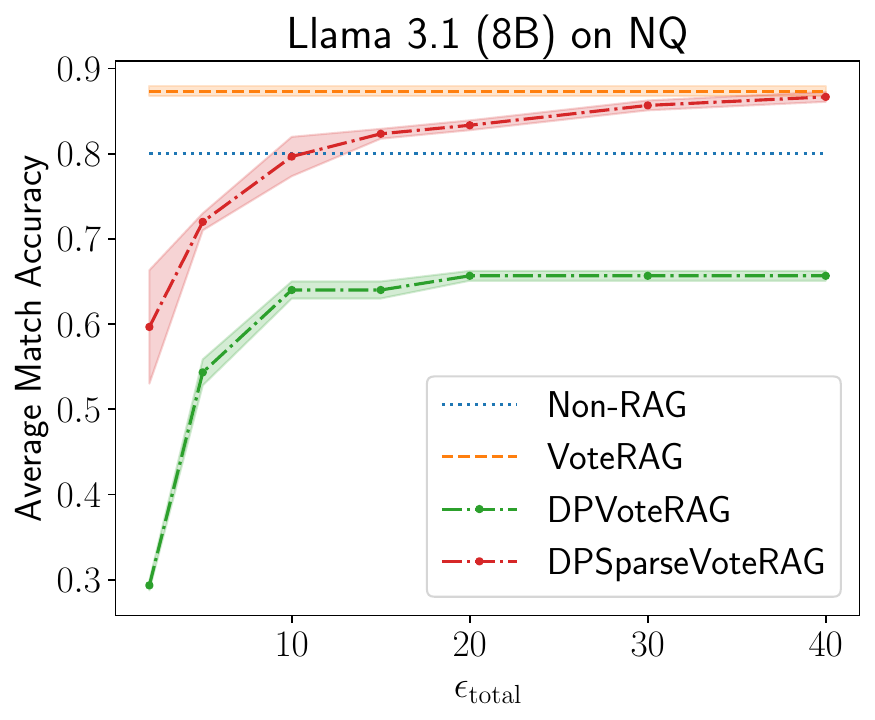}
    
    \includegraphics[width=0.3\textwidth]{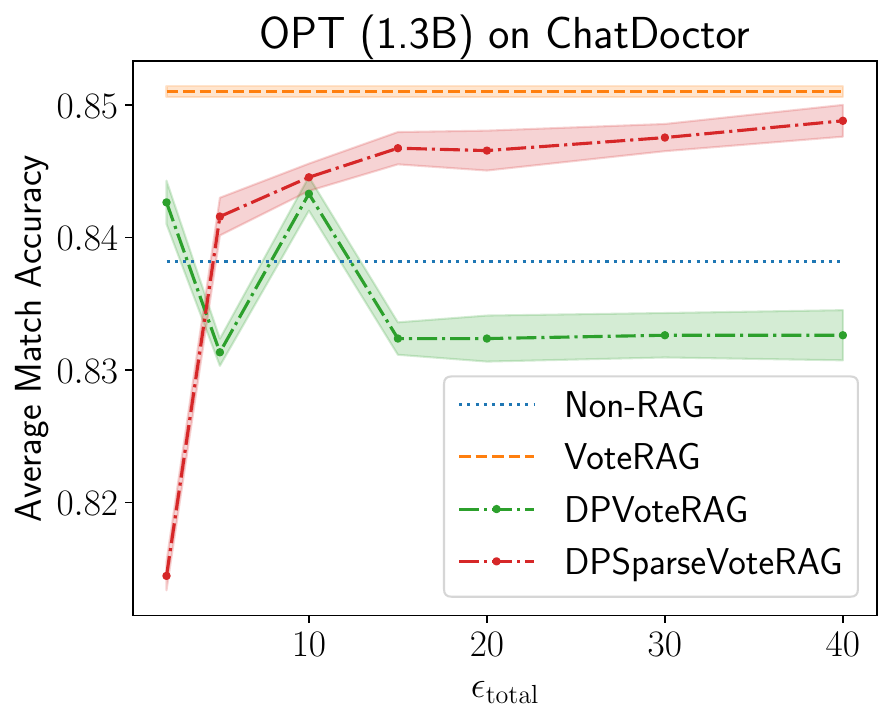}
    \hfill
    \includegraphics[width=0.3\textwidth]{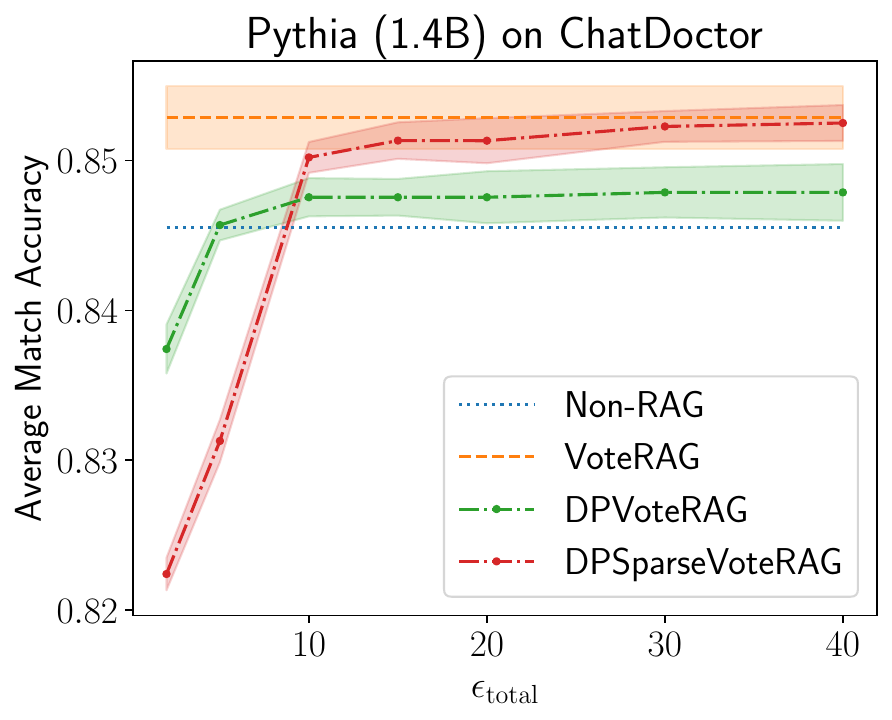}
    \hfill
    \includegraphics[width=0.3\textwidth]{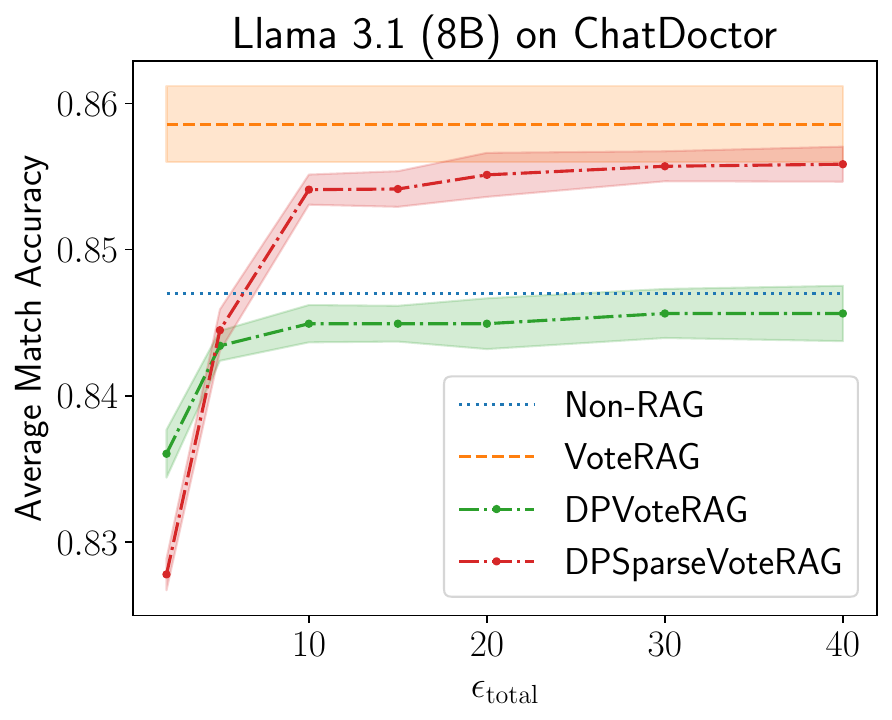}
    
    \caption{Average match accuracy comparison across algorithms on Trivia (upper row) and NQ (lower row) datasets with different generator LLMs: OPT (1.3B) (left column), Pythia (1.4B) (middle column), and Llama 3.1 (8B) (right column).
    The reported results are the means and standard deviations of average match accuracy over three runs.
    We report the best results over hyperparameters for each $\epsilon_\mathrm{total}$.}
    \label{fig:main-result}
\end{figure*}

\begin{table*}[t]
    \caption{Average match accuracy comparison of our algorithms on Trivia dataset with OPT (1.3B) under varying values of total privacy budget $\epsilon_\mathrm{total}$ with different hyperparameters, the number of voters $m$ and $\epsilon_\mathrm{token}$. Bold font values represent the best performance of each algorithm under fixed $\epsilon_\mathrm{total}$.
    We report the means of average match accuracy over three runs.}
    \label{tab:detailed-result-trivia-opt1.3b}
    \centering
    \begin{tabular}{l c *{4}{c c c}}
    \toprule
    Algorithm & 
    & \multicolumn{3}{c}{$\epsilon_\mathrm{total} = 5$} & \multicolumn{3}{c}{$\epsilon_\mathrm{total} = 10$} & 
    \multicolumn{3}{c}{$\epsilon_\mathrm{total} = 20$} & \multicolumn{3}{c}{$\epsilon_\mathrm{total} = 40$} \\
    \cmidrule(lr){3-5} \cmidrule(lr){6-8} \cmidrule(lr){9-11} \cmidrule(lr){12-14}
    & $m$ & 30 & 40 & 50 & 30 & 40 & 50 & 30 & 40 & 50 & 30 & 40 & 50 \\
    \midrule
    \textbf{\dpvoterag{}} & & & & & & & & & & & & & \\
    $\epsilon_\mathrm{token}=1$ & & 
    $0.37$&$\bm{0.40}$&$\bm{0.40}$& $0.45$&$\bm{0.49}$&$\bm{0.49}$& $0.45$&$0.49$&$0.49$& $0.45$&$0.49$&$0.49$ \\
    $\epsilon_\mathrm{token}=2$ & & 
    $0.25$&$0.27$&$0.27$& $0.42$&$0.41$&$0.41$& $\bm{0.51}$&$0.50$&$0.49$& $\bm{0.51}$&$0.50$&$0.49$ \\
    $\epsilon_\mathrm{token}=5$ & &
    $0.12$&$0.12$&$0.12$& $0.25$&$0.27$&$0.27$& $0.38$&$0.38$&$0.38$& $0.46$&$0.46$&$0.44$ \\
    \midrule
    \textbf{\dpsparsevoterag{}} & & & & & & & & & & & & & \\
    $\epsilon_\mathrm{token}=1$ & & 
    $0.28$&$0.37$&$0.45$& $0.28$&$0.38$&$0.46$& $0.28$&$0.38$&$0.46$& $0.28$&$0.38$&$0.46$ \\
    $\epsilon_\mathrm{token}=2$ & & 
    $0.42$&$0.45$&$\bm{0.48}$& $0.48$&$0.55$&$\bm{0.57}$& $0.49$&$0.57$&$\bm{0.60}$& $0.49$&$0.57$&$0.60$ \\
    $\epsilon_\mathrm{token}=5$ & & 
    $0.33$&$0.33$&$0.34$& $0.47$&$0.48$&$0.49$& $0.56$&$0.55$&$0.55$& $0.59$&$0.60$&$\bm{0.61}$ \\
    \bottomrule
    \end{tabular}
\end{table*}
\paragraph{Our RAG algorithms boost the QA accuracy even under a formal privacy guarantee.}
Figure~\ref{fig:main-result} shows the average match accuracy of baseline algorithms and our private algorithms under different total privacy guarantees ($\epsilon_\mathrm{total}$).
Across different datasets and LLMs, we observe that \dpsparsevoterag{} outperforms \nonrag{} mostly under $\epsilon_\mathrm{total} \geq 10$ and approaches the upper bound of \voterag{} as we allow a larger privacy budget.
This demonstrates that our algorithms enable us to exploit the external knowledge through RAG to improve the utility of QA tasks \emph{while} ensuring a reasonable level of privacy.

\paragraph{\dpsparsevoterag{} is strictly better than \dpvoterag{}.}
In Figure~\ref{fig:main-result}, we find that \dpsparsevoterag{} consistently outperforms \dpvoterag{} across different LLMs and datasets. \dpsparsevoterag{} augments \dpvoterag{} by utilizing the non-RAG LLM and the sparse vector technique so that it only spends a privacy budget for an output token requiring sensitive external knowledge. The consistently better performances of \dpsparsevoterag{} suggest the importance of separately treating token generations for meaningful tokens, i.e., tokens requiring external knowledge, and for other general tokens in the privacy-constraint setting.

\paragraph{$\epsilon_\mathrm{token}$ should allow medium-length outputs. $m$ should balance the DP noise and \# of well-informed voters.}
We take a closer look at the effects of the hyperparameters in Table~\ref{tab:detailed-result-trivia-opt1.3b} with OPT (1.3B) on Trivia dataset under different total privacy budgets $\epsilon_\mathrm{total}$. 
We provide the detailed results, as in Table~\ref{tab:detailed-result-trivia-opt1.3b}, with other LLMs in Appendix~\ref{appendix:additional-exp}.

Commonly between our private algorithms, we observe that the optimal $\epsilon_\mathrm{token}$ increases as we allow more total privacy budgets. 
Under a tight total privacy budget, large $\epsilon_\mathrm{token}$ allows our algorithms to only output a few meaningful tokens; thus, smaller $\epsilon_\mathrm{token}$ is preferable.
Conversely, under a large total privacy budget, accurate token generation with large $\epsilon_\mathrm{token}$ is more important than having more tokens generated with small $\epsilon_\mathrm{token}$.
Therefore, it is advised that we set $\epsilon_\mathrm{token}$ to be as large as possible to enable accurate token generations \emph{as long as} it is small enough to allow the algorithms to generate a reasonably large number of tokens ($\approx 10$).
Notice that \dpsparsevoterag{} generally allows us to set larger $\epsilon_\mathrm{token}$ than \dpvoterag{} under a fixed total privacy budget. This implies the benefit of \dpsparsevoterag{} to save and spend a privacy budget cleverly---it can spend the saved privacy budget for generating important tokens for answering questions correctly.

With regard to the number of voters $m$, we generally see that more voters yield better utility with $\epsilon_\mathrm{token}=1$, but the number of voters has less effect on the utility with larger $\epsilon_\mathrm{token}$.
This is due to the two distinct consequences of having more voters.
More voters alleviate the effect of DP noise on the token histograms constructed in the algorithms. 
However, depending on the number of relevant documents to the question, there is a risk of having voters with irrelevant documents who can vote for the wrong tokens.
The first consequence is more dominant particularly under small $\epsilon_\mathrm{token}$ while the second is more dominant under larger $\epsilon_\mathrm{token}$.
Hence, $m$ should be set to balance these two consequences for achieving better per-token generation quality.

\subsection{Empirical Privacy Evaluation}
\begin{figure*}[!t]
    \centering
    \includegraphics[width=0.3\textwidth]{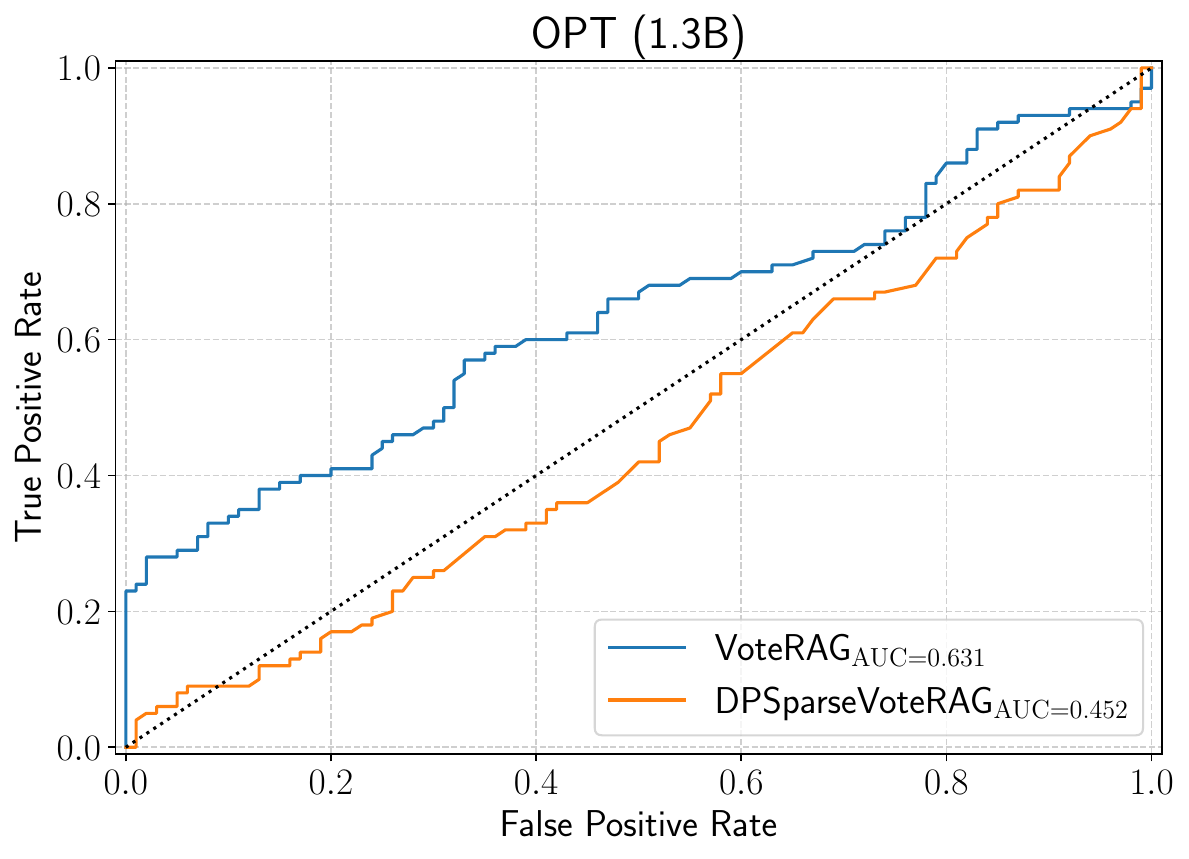}
    \hfill
    \includegraphics[width=0.3\textwidth]{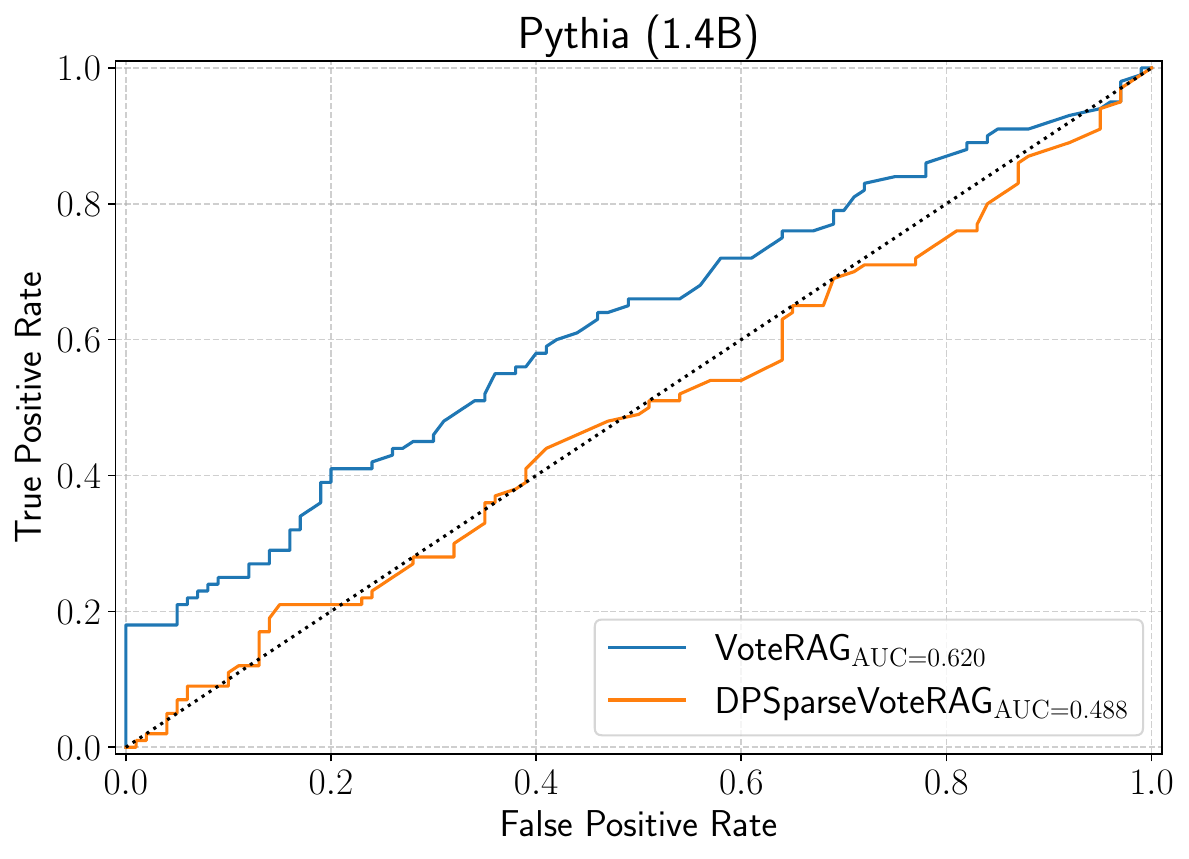}
    \hfill
    \includegraphics[width=0.3\textwidth]{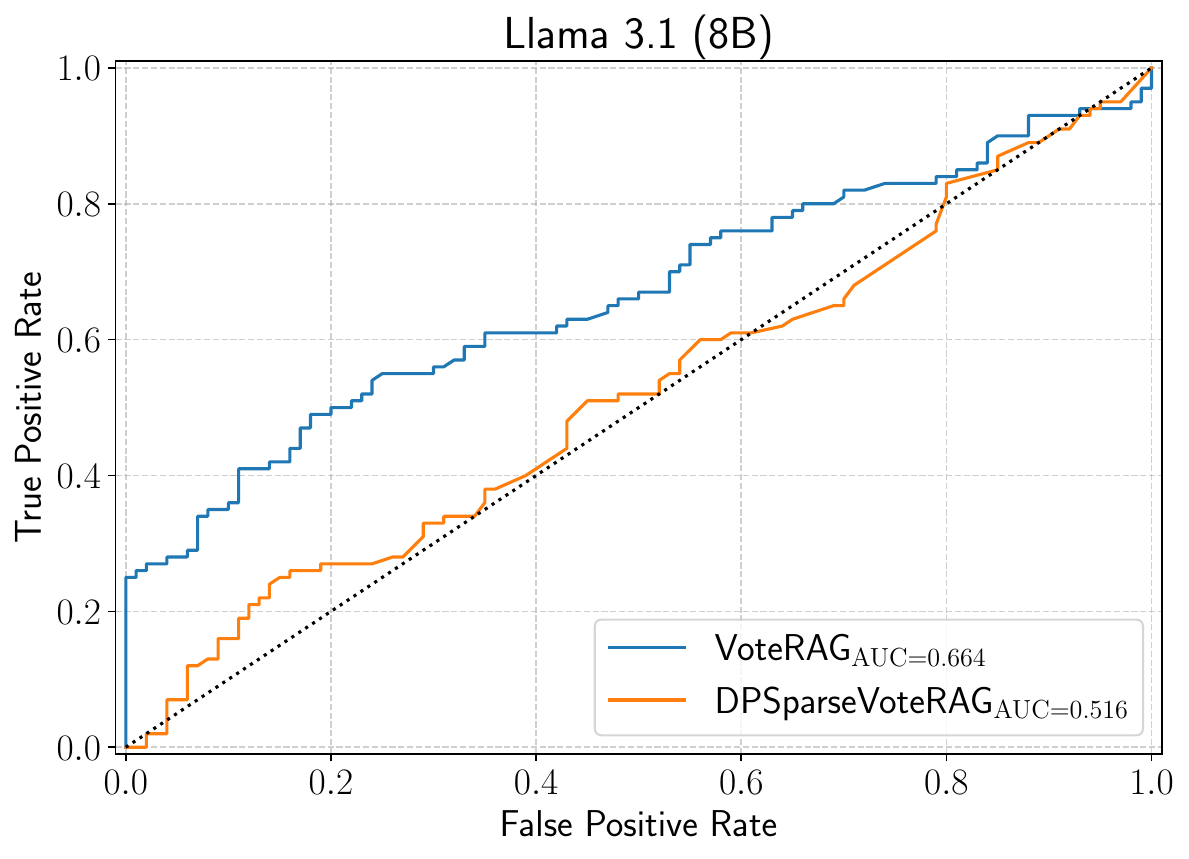}
    
    \caption{TPR-FPR curve of S$^2$MIA for VoteRAG and DPSparseVoteRAG ($\varepsilon$=10) when the base LLMs are OPT (1.3B), Pythia (1.4B) and Llama 3.1 (8B).
    }
    \label{fig:attack}
\end{figure*}

To assess the degree of privacy protection offered by our proposed method, we evaluate the vulnerability of both a non-private RAG system and our privacy-preserving RAG system on the privacy-sensitive ChatDoctor dataset using membership inference attacks (MIA). Given a target document $x$ and a system $f_D$, an MIA computes a score $s(x, f)$ that reflects the likelihood of $x \in D$. Without loss of generality, we assume higher scores indicate a greater probability of membership. By applying the attack to two sets of documents (an in-distribution set $D_{\text{in}} \subset D$ and an out-of-distribution set $D_{\text{out}}$ with no overlap with $D$), we can derive a TPR–FPR curve and compute its AUC.

We adopt the membership score design from S$^2$MIA~\citep{li2025generating}. In the ChatDoctor dataset, each document corresponds to a patient–doctor conversation. For a target document $x$, we extract the patient’s query $x_t^q$ and measure the similarity between the response $x_t^r$ generated by the tested RAG system and the doctor’s ground-truth answer $x_t^g$ in $x$. The similarity is quantified using the average precision score defined in BLEU~\citep{papineni2002bleu}, which serves directly as the membership score in S$^2$MIA.

Figure~\ref{fig:attack} presents the TPR–FPR curves and the corresponding AUC values. Without any privacy protection (VoteRAG), the attack is highly effective, yielding AUC values well above the diagonal baseline (0.5). In contrast, when querying our privacy-preserving system DPSparseVoteRAG with $\varepsilon = 10$, the attack performance collapses to the naive baseline (AUC $\approx 0.5$). This demonstrates that our method effectively mitigates empirical privacy attacks while maintaining strong utility on the QA task, as shown in Figure~\ref{fig:main-result}.

\subsection{More Analysis of DPSparseVoteRAG and DPVoteRAG}
\begin{figure*}[!t]
    \centering
    \includegraphics[width=0.9\textwidth]{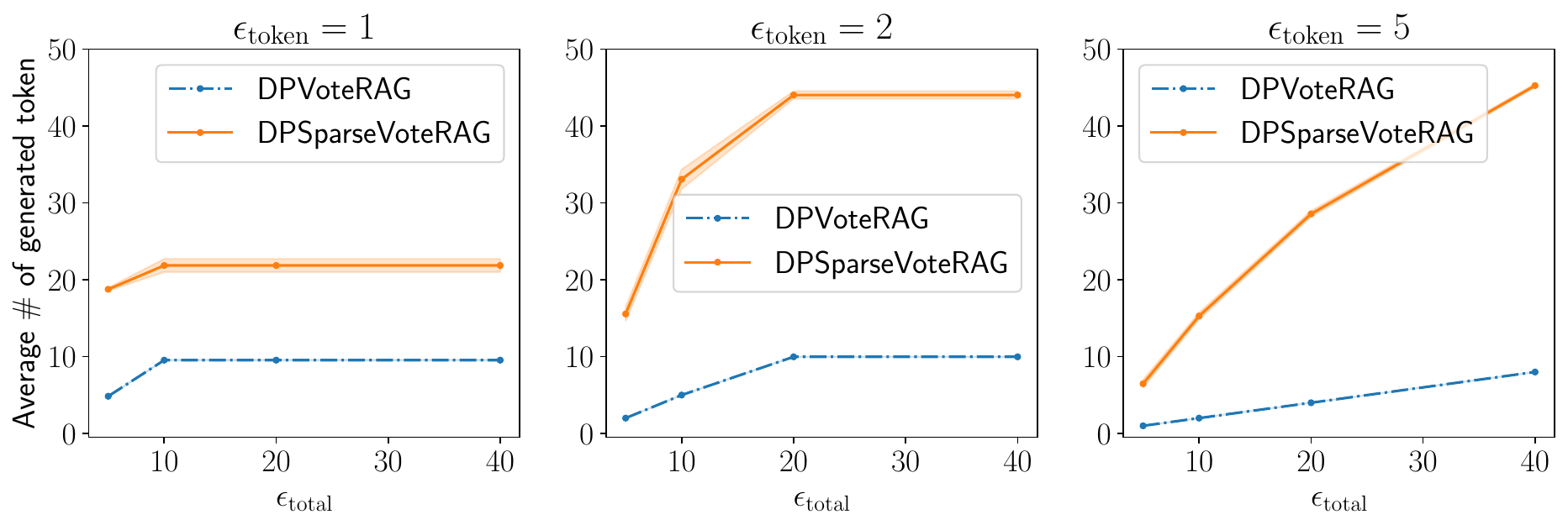}
    
    \caption{Average numbers of generated tokens for each $\epsilon_\mathrm{token}$ and $\epsilon_\mathrm{total}$ with OPT (1.3B) on Trivia dataset. We fix $m=50$.
    We report the means and standard deviations over three runs. We see \dpsparsevoterag{} generates much more tokens than \dpvoterag{}.}
    \label{fig:num-generated-tokens}
\end{figure*}

\paragraph{The length of generation.} The design of DPSparseVoteRAG might allow longer generation than DPVoteRAG because of the tighter composition from SVT. We empirically validated this intuition.
Figure~\ref{fig:num-generated-tokens} shows the numbers of tokens generated by \dpvoterag{} and \dpsparsevoterag{}. As we expect by the design of \dpsparsevoterag{}, we see \dpsparsevoterag{} generates much more tokens than \dpvoterag{}. This implies the effectiveness of the sparse vector technique in \dpsparsevoterag{} to smartly spend privacy budget enabling long enough token sequences.

\paragraph{Effects of number of ground truth relevant documents.}
\begin{figure}[!t]
    \centering
    \includegraphics[width=0.45\textwidth]{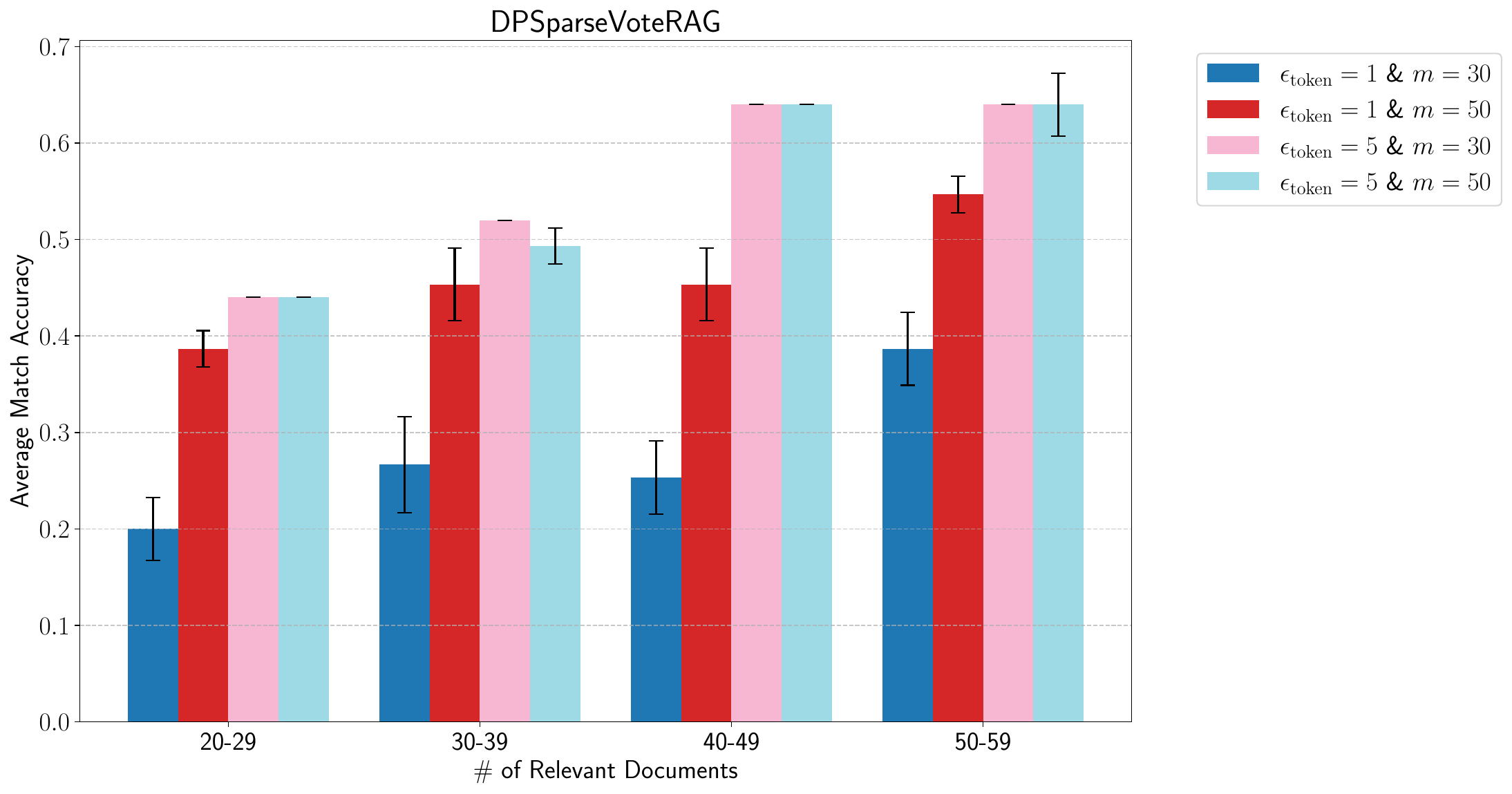}
    
    \caption{Average match accuracy for questions with different numbers of ground truth relevant documents across baseline algorithms and our algorithms with OPT (1.3B) on Trivia dataset.
    We report the means and standard deviations over three runs. We see questions with more relevant documents tend to be answered correctly by our algorithm.}
    \label{fig:num-relevant-docs}
\end{figure}
Figure~\ref{fig:num-relevant-docs} shows the performances for different numbers of ground truth relevant documents.
We see questions with more relevant documents tend to be answered correctly by our algorithm.

%% file: related_work.tex
\section{Related Work}
\paragraph{Privacy-preserving algorithms in large language models.}
\citet{zeng-etal-2024-good} proposed an empirical privacy-preserving algorithm for RAG through the synthetic data generation, while our work studies privacy-preserving RAG in the framework differential privacy, which protects the privacy of each individual document with the theoretical guarantee.
Differential privacy has been studied in many other tasks in large language models too. 
\textit{Prompt tuning} helps tailor the LLM to new tasks from a (private) test-domain dataset.
\citet{hong2024dpopt} and \citet{duan2024flocks} study the DP mechanism on two different prompt tuning frameworks: prompt optimization and offsite prompt tuning~\citep{sordoni2023joint}.
\textit{In-context learning} adapts to different tasks by illustrating some examples in the context as the task description.
DP in-context learning considers the situation when the examples are picked from any private set.
\citet{tang2024privacypreserving} tackles this problem by generating synthetic examples with DP and \citet{wu2024privacypreserving} solves the DP test query by generating the answers, both in a sample-and-aggregate fashion.
\citet{amin2024private} proposes the aggregation based method to generate synthetic texts with DP, which applies the similar SVT idea of our methodology to save the budget for some tokens.
The differentially private \textit{pretraining and finetuning} of LLMs has been studied to address the privacy concern in the training data and memory is a large bottleneck when naively deploying DP-SGD~\citep{abadi2016deep}.
\citet{li2022large} focuses on the pretraining stage which introduces ghost clipping to make DP-SGD more memory efficient.
\citet{yu2022differentially} explores finetuning in the parameter-efficient framework LoRA~\citep{hu2022lora}.
Notice that DP voting plays a crucial role in these sample-and-aggregate algorithms, including ours.
A basic approach is to apply the Laplacian or Gaussian mechanism~\citep{dwork_algorithmic_2014}.
\citet{papernot_semi-supervised_2017-1,papernot_scalable_2018-2} proposed a data-dependent privacy analysis, which can be tighter when the majority vote has a large margin over other options.
We integrate the LimitedDomain mechanism for our algorithm, which addresses challenges when the voting domain is large~\citep{durfee_practical_2019-1}; the large vocabulary size in token voting is our main bottleneck.

\paragraph{Composition in differential privacy.} Our algorithms generate the answers token by token, where each token needs a query to the private dataset and consumes some privacy budget.
In this paper, we set up the privacy parameters before the start of the algorithm and have a pre-set maximum number of tokens to generate.
However, the number of tokens to generate is different per question and is unknown before the algorithm starts -- it is possible that the number of generated tokens is much smaller than the pre-set number but we still need to pay the full pre-defined privacy cost.
A line of work~\citep{rogers2016privacy, feldman2021individual, lecuyer2021practical, whitehouse2023fully} tries to measure the privacy budget in fully adaptive composition where the budget consuming can interact with the data.
Especially, \citet{whitehouse2023fully} gives an analysis for this fully adaptive setting which matches the tightness of advanced composition.
The idea of fully adaptive composition sounds a fit to our problem, which allows us to ``pay as we go'', rather than predefining the $\epsilon_{\mathrm{total}}$ before the generation process.
We found the analysis for fully adaptive setting is effective for large number of steps and small budget per step, while in our algorithm the number of generated tokens would not be very large and each token generation needs a relatively large budget to guarantee the utility.
This mismatch makes us stick with the advanced composition.

%% file: conclusion.tex
\section{Conclusion and Future Work}
We introduce the first differentially private algorithms for RAG, enabling us to enhance LLMs by domain-specific but sensitive external corpus.
With our novel combination of the DP voting algorithm and sparse vector technique along with the non-private LLM, we succeed in spending privacy budget only when the LLM needs sensitive information to generate a new token. Consequently, \dpsparsevoterag{} generates a sufficiently long and accurate response under a reasonable privacy budget.
Our experiments demonstrate that our algorithms outperform the non-RAG baseline across different datasets and models, showing their effectiveness.

One of our future directions is to conduct more practical empirical evaluations. The Wikipedia dataset, which we use as the external data source, is typically included in the training data of recent LLMs. RAG is particularly effective when the external knowledge is \emph{truly sensitive} and thus outside the LLM training data. 
It is essential for us to conduct evaluations that are as close to the real situation as possible and see how effective our algorithms are over non-RAG LLMs.
Since our usage of the sparse vector technique is applicable to any DP token generation algorithm through voting, another future direction would be to examine how it improves DP token generation across different tasks, e.g., in-context learning and prompt tuning.

%% file: appendix.tex
%
\subsection{Additional Experimental Results}
\label{appendix:additional-exp}
\begin{figure*}[h]
    \centering
    \includegraphics[width=0.4\textwidth]{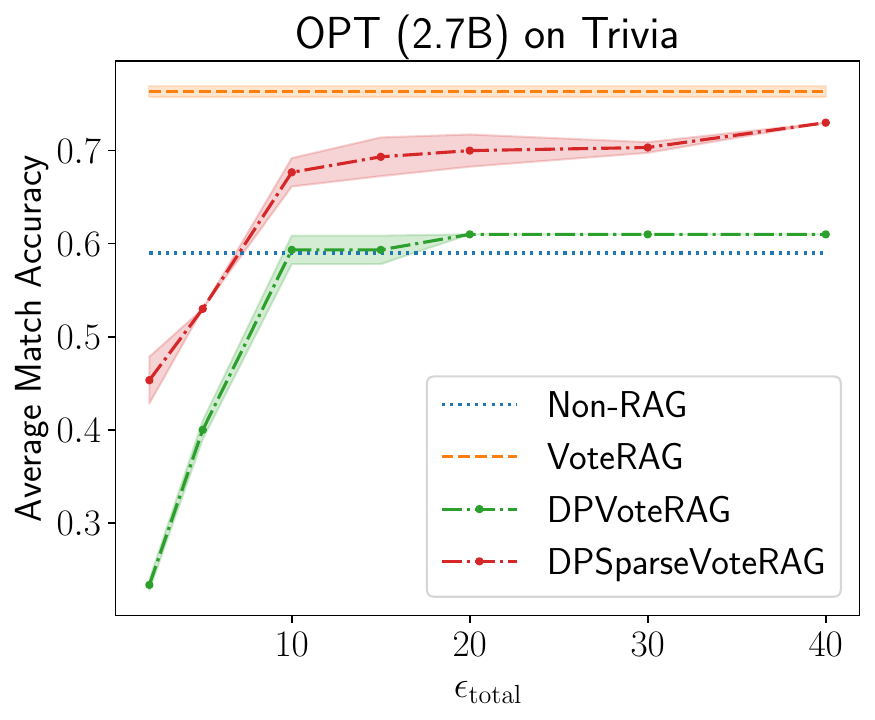}
    \hfill
    \includegraphics[width=0.4\textwidth]{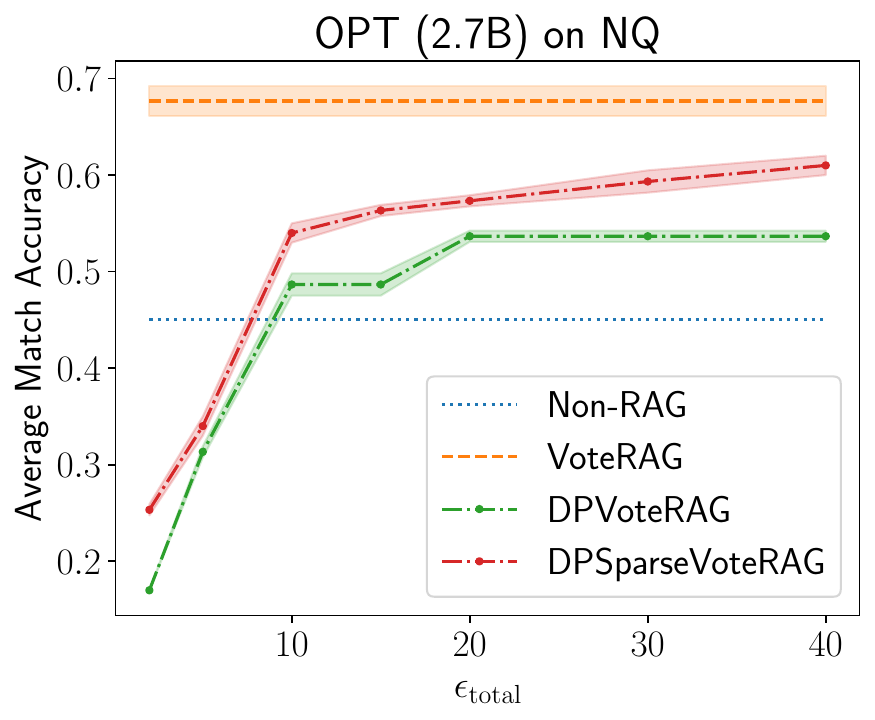}
    
    \caption{Average match accuracy comparison across algorithms on Trivia (left) and NQ (right) datasets with OPT (2.7B).
    The reported results are the means and standard deviations of average match accuracy over three runs.
    We report the best results over hyperparameters for each $\epsilon_\mathrm{total}$.}
    \label{fig:main-result-opt27}
\end{figure*}

\begin{figure*}[h]
    \centering
    \includegraphics[width=0.4\textwidth]{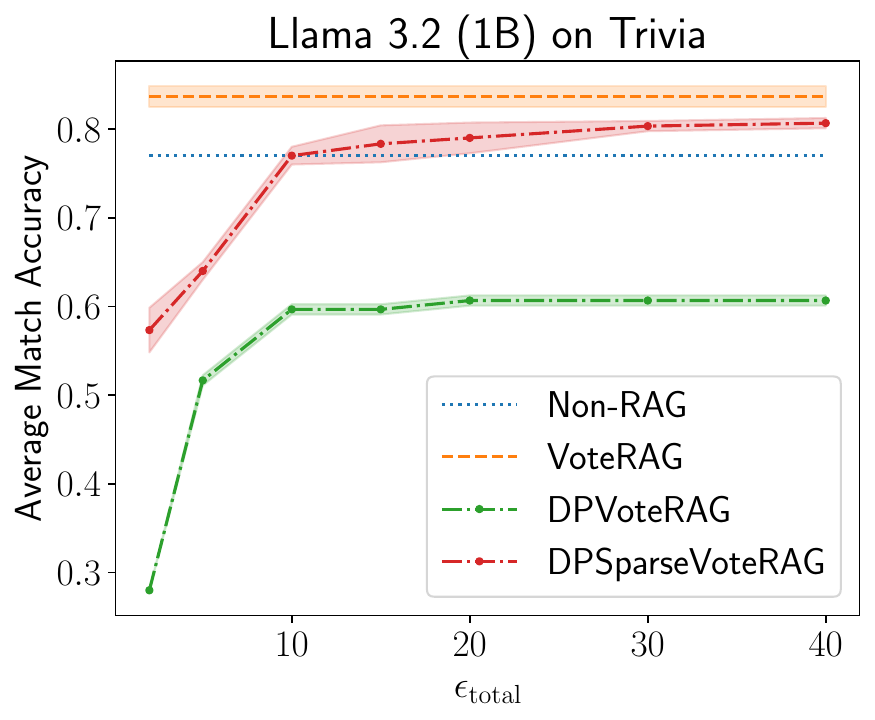}
    \hfill
    \includegraphics[width=0.4\textwidth]{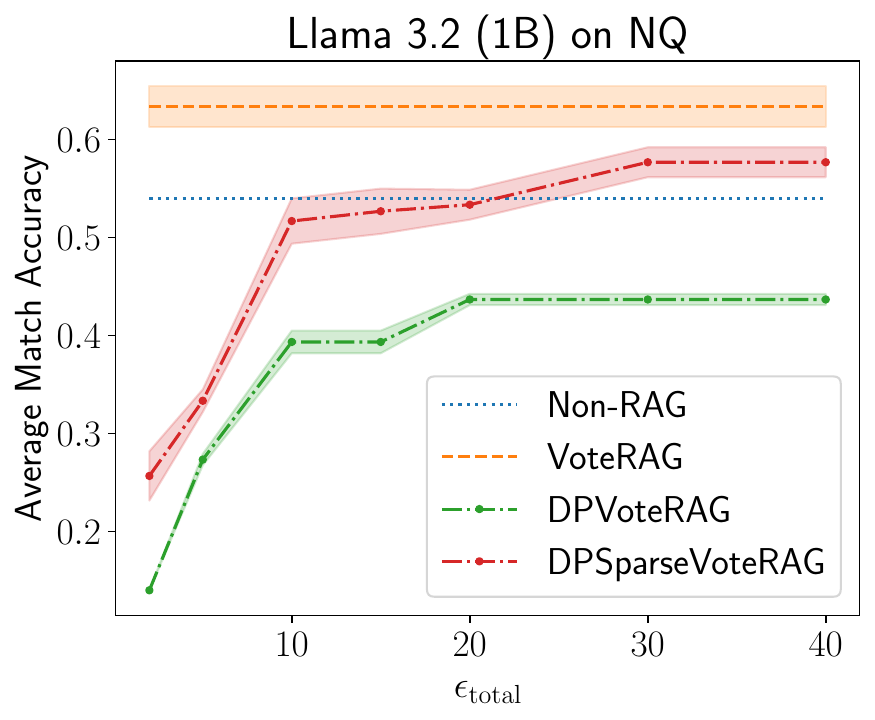}
    
    \caption{Average match accuracy comparison across algorithms on Trivia (left) and NQ (right) datasets with Llama 3.2 (1B).
    The reported results are the means and standard deviations of average match accuracy over three runs.
    We report the best results over hyperparameters for each $\epsilon_\mathrm{total}$.}
    \label{fig:main-result-llama32}
\end{figure*}

\begin{figure*}[h]
    \centering
    \includegraphics[width=0.4\textwidth]{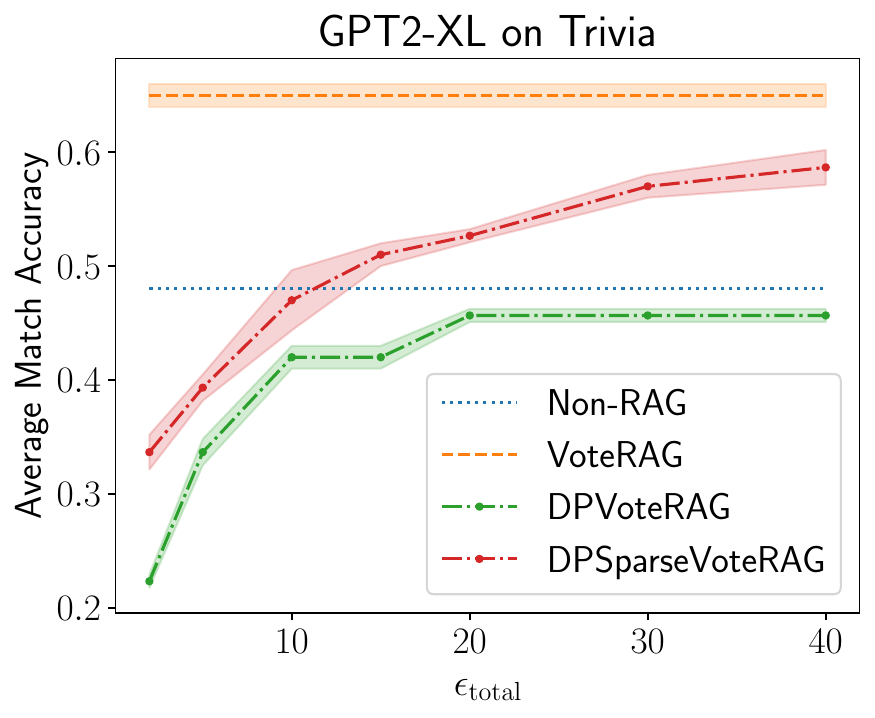}
    \hfill
    \includegraphics[width=0.4\textwidth]{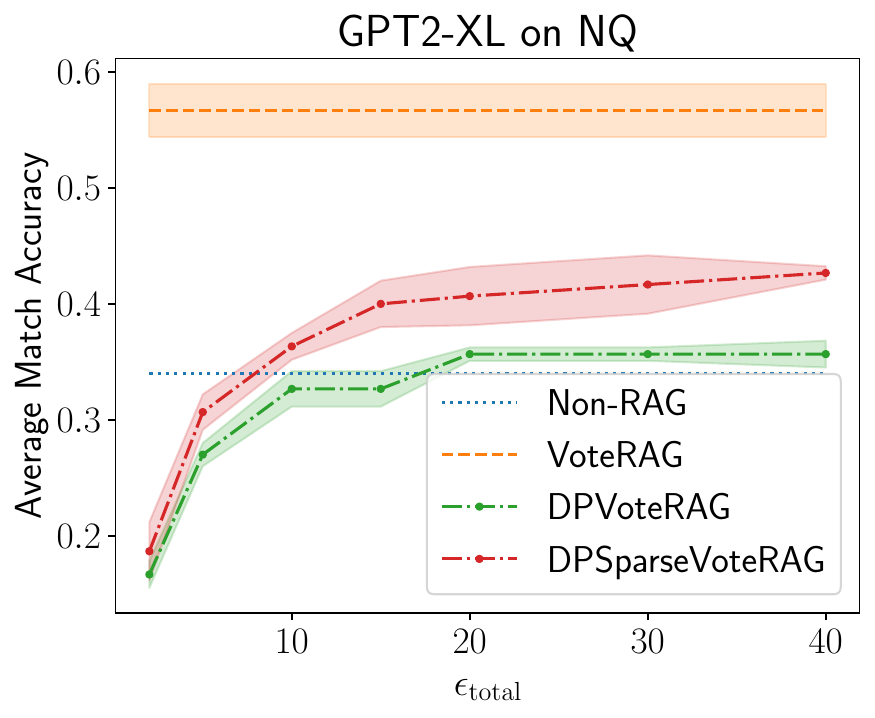}
    
    \caption{Average match accuracy comparison across algorithms on Trivia (left) and NQ (right) datasets with GPT2-XL.
    The reported results are the means and standard deviations of average match accuracy over three runs.
    We report the best results over hyperparameters for each $\epsilon_\mathrm{total}$.}
    \label{fig:main-result-gpt2-xl}
\end{figure*}

In Figures~\ref{fig:main-result-opt27}--\ref{fig:main-result-gpt2-xl}, we present the average match accuracy of baseline algorithms and our algorithms for different total privacy guarantees ($\epsilon_\mathrm{total}$) with OPT (2.7B), Llama 3.2 (1B), and GPT2-XL. 
We see the similar trend observed in Figure~\ref{fig:main-result}.

For completeness, we further present the detailed results, in the same form of the one provided in Table~\ref{tab:detailed-result-trivia-opt1.3b}, with OPT (1.3B and 2.7B), Pythia (1.4B), Llama 3.1 (8B), Llama 3.2 (1B), and GPT2-XL on Trivia and NQ datasets in Tables~\ref{tab:detailed-result-nq-opt1.3b}--\ref{tab:detailed-result-nq-gpt2}.

\begin{table*}[htbp]
    \caption{Average match accuracy comparison of our algorithms on NQ dataset with OPT (1.3B) under varying values of total privacy budget $\epsilon_\mathrm{total}$ with different hyperparameters, the number of voters $m$ and $\epsilon_\mathrm{token}$. Bold font values represent the best performance of each algorithm under fixed $\epsilon_\mathrm{total}$.
    We report the means of average match accuracy over three runs.}
    \label{tab:detailed-result-nq-opt1.3b}
    \centering
    \begin{tabular}{l c *{4}{c c c}}
    \toprule
    Algorithm & 
    & \multicolumn{3}{c}{$\epsilon_\mathrm{total} = 5$} & \multicolumn{3}{c}{$\epsilon_\mathrm{total} = 10$} & 
    \multicolumn{3}{c}{$\epsilon_\mathrm{total} = 20$} & \multicolumn{3}{c}{$\epsilon_\mathrm{total} = 40$} \\
    \cmidrule(lr){3-5} \cmidrule(lr){6-8} \cmidrule(lr){9-11} \cmidrule(lr){12-14}
    & $m$ & 30 & 40 & 50 & 30 & 40 & 50 & 30 & 40 & 50 & 30 & 40 & 50 \\
    \midrule
    \textbf{\dpvoterag{}} & & & & & & & & & & & & & \\
    $\epsilon_\mathrm{token}=1$ & & 
    $0.17$&$\bm{0.23}$&$\bm{0.23}$& $0.28$&$0.35$&$\bm{0.37}$& $0.28$&$0.35$&$0.37$& $0.28$&$0.35$&$0.37$ \\
    $\epsilon_\mathrm{token}=2$ & & 
    $0.12$&$0.14$&$0.13$& $0.24$&$0.25$&$0.26$& $0.38$&$\bm{0.42}$&$\bm{0.42}$& $0.38$&$\bm{0.42}$&$\bm{0.42}$ \\
    $\epsilon_\mathrm{token}=5$ & &
    $0.04$&$0.06$&$0.05$& $0.12$&$0.14$&$0.13$& $0.22$&$0.22$&$0.21$& $0.36$&$0.36$&$0.35$ \\
    \midrule
    \textbf{\dpsparsevoterag{}} & & & & & & & & & & & & & \\
    $\epsilon_\mathrm{token}=1$ & & 
    $0.14$&$0.18$&$0.28$& $0.14$&$0.18$&$0.29$& $0.14$&$0.18$&$0.29$& $0.14$&$0.18$&$0.29$ \\
    $\epsilon_\mathrm{token}=2$ & & 
    $0.23$&$0.25$&$\bm{0.29}$& $0.37$&$0.41$&$\bm{0.43}$& $0.39$&$0.44$&$\bm{0.47}$& $0.39$&$0.44$&$0.47$ \\
    $\epsilon_\mathrm{token}=5$ & & 
    $0.11$&$0.14$&$0.13$& $0.28$&$0.31$&$0.29$& $0.43$&$0.43$&$0.42$& $\bm{0.53}$&$\bm{0.53}$&$\bm{0.53}$ \\
    \bottomrule
    \end{tabular}
\end{table*}

\begin{table*}[htbp]
    \caption{Average match accuracy comparison of our algorithms on Trivia dataset with OPT (2.7B) under varying values of total privacy budget $\epsilon_\mathrm{total}$ with different hyperparameters, the number of voters $m$ and $\epsilon_\mathrm{token}$. Bold font values represent the best performance of each algorithm under fixed $\epsilon_\mathrm{total}$.
    We report the means of average match accuracy over three runs.}
    \label{tab:detailed-result-trivia-opt2.7b}
    \centering
    \begin{tabular}{l c *{4}{c c c}}
    \toprule
    Algorithm & 
    & \multicolumn{3}{c}{$\epsilon_\mathrm{total} = 5$} & \multicolumn{3}{c}{$\epsilon_\mathrm{total} = 10$} & 
    \multicolumn{3}{c}{$\epsilon_\mathrm{total} = 20$} & \multicolumn{3}{c}{$\epsilon_\mathrm{total} = 40$} \\
    \cmidrule(lr){3-5} \cmidrule(lr){6-8} \cmidrule(lr){9-11} \cmidrule(lr){12-14}
    & $m$ & 30 & 40 & 50 & 30 & 40 & 50 & 30 & 40 & 50 & 30 & 40 & 50 \\
    \midrule
    \textbf{\dpvoterag{}} & & & & & & & & & & & & & \\
    $\epsilon_\mathrm{token}=1$ & & 
    $0.34$&$\bm{0.40}$&$0.39$& $0.48$&$0.58$&$\bm{0.59}$& $0.48$&$0.58$&$0.59$& $0.48$&$0.58$&$0.59$ \\
    $\epsilon_\mathrm{token}=2$ & & 
    $0.24$&$0.23$&$0.23$& $0.41$&$0.41$&$0.40$& $0.59$&$\bm{0.61}$&$0.60$& $0.59$&$\bm{0.61}$&$0.60$ \\
    $\epsilon_\mathrm{token}=5$ & &
    $0.09$&$0.10$&$0.10$& $0.23$&$0.24$&$0.23$& $0.38$&$0.39$&$0.38$& $0.49$&$0.51$&$0.51$ \\
    \midrule
    \textbf{\dpsparsevoterag{}} & & & & & & & & & & & & & \\
    $\epsilon_\mathrm{token}=1$ & & 
    $0.31$&$0.48$&$\bm{0.53}$& $0.31$&$0.49$&$0.53$& $0.31$&$0.49$&$0.53$& $0.31$&$0.49$&$0.53$ \\
    $\epsilon_\mathrm{token}=2$ & & 
    $0.50$&$0.52$&$\bm{0.53}$& $0.57$&$0.66$&$\bm{0.68}$& $0.58$&$0.69$&$\bm{0.70}$& $0.58$&$0.69$&$0.70$ \\
    $\epsilon_\mathrm{token}=5$ & & 
    $0.35$&$0.38$&$0.37$& $0.54$&$0.54$&$0.54$& $0.66$&$0.66$&$0.66$& $\bm{0.73}$&$0.71$&$0.70$ \\
    \bottomrule
    \end{tabular}
\end{table*}

\begin{table*}[htbp]
    \caption{Average match accuracy comparison of our algorithms on NQ dataset with OPT (2.7B) under varying values of total privacy budget $\epsilon_\mathrm{total}$ with different hyperparameters, the number of voters $m$ and $\epsilon_\mathrm{token}$. Bold font values represent the best performance of each algorithm under fixed $\epsilon_\mathrm{total}$.
    We report the means of average match accuracy over three runs.}
    \label{tab:detailed-result-nq-opt2.7b}
    \centering
    \begin{tabular}{l c *{4}{c c c}}
    \toprule
    Algorithm & 
    & \multicolumn{3}{c}{$\epsilon_\mathrm{total} = 5$} & \multicolumn{3}{c}{$\epsilon_\mathrm{total} = 10$} & 
    \multicolumn{3}{c}{$\epsilon_\mathrm{total} = 20$} & \multicolumn{3}{c}{$\epsilon_\mathrm{total} = 40$} \\
    \cmidrule(lr){3-5} \cmidrule(lr){6-8} \cmidrule(lr){9-11} \cmidrule(lr){12-14}
    & $m$ & 30 & 40 & 50 & 30 & 40 & 50 & 30 & 40 & 50 & 30 & 40 & 50 \\
    \midrule
    \textbf{\dpvoterag{}} & & & & & & & & & & & & & \\
    $\epsilon_\mathrm{token}=1$ & & 
    $0.26$&$0.30$&$\bm{0.31}$& $0.39$&$0.48$&$\bm{0.49}$& $0.39$&$0.48$&$0.49$& $0.39$&$0.48$&$0.49$ \\
    $\epsilon_\mathrm{token}=2$ & & 
    $0.20$&$0.18$&$0.17$& $0.36$&$0.34$&$0.33$& $\bm{0.54}$&$0.51$&$0.51$& $\bm{0.54}$&$0.51$&$0.51$ \\
    $\epsilon_\mathrm{token}=5$ & &
    $0.06$&$0.05$&$0.05$& $0.20$&$0.17$&$0.17$& $0.34$&$0.31$&$0.30$& $0.46$&$0.43$&$0.42$ \\
    \midrule
    \textbf{\dpsparsevoterag{}} & & & & & & & & & & & & & \\
    $\epsilon_\mathrm{token}=1$ & & 
    $0.16$&$0.24$&$0.32$& $0.16$&$0.24$&$0.32$& $0.16$&$0.24$&$0.32$& $0.16$&$0.24$&$0.32$ \\
    $\epsilon_\mathrm{token}=2$ & & 
    $0.31$&$0.32$&$\bm{0.34}$& $0.46$&$0.52$&$\bm{0.54}$& $0.47$&$0.53$&$\bm{0.57}$& $0.47$&$0.53$&$0.57$ \\
    $\epsilon_\mathrm{token}=5$ & & 
    $0.12$&$0.12$&$0.11$& $0.38$&$0.35$&$0.35$& $\bm{0.57}$&$0.53$&$0.53$& $\bm{0.61}$&$0.58$&$0.60$ \\
    \bottomrule
    \end{tabular}
\end{table*}

\begin{table*}[htbp]
    \caption{Average match accuracy comparison of our algorithms on Trivia dataset with Pythia (1.4B) under varying values of total privacy budget $\epsilon_\mathrm{total}$ with different hyperparameters, the number of voters $m$ and $\epsilon_\mathrm{token}$. Bold font values represent the best performance of each algorithm under fixed $\epsilon_\mathrm{total}$.
    We report the means of average match accuracy over three runs.}
    \label{tab:detailed-result-trivia-pythia}
    \centering
    \begin{tabular}{l c *{4}{c c c}}
    \toprule
    Algorithm & 
    & \multicolumn{3}{c}{$\epsilon_\mathrm{total} = 5$} & \multicolumn{3}{c}{$\epsilon_\mathrm{total} = 10$} & 
    \multicolumn{3}{c}{$\epsilon_\mathrm{total} = 20$} & \multicolumn{3}{c}{$\epsilon_\mathrm{total} = 40$} \\
    \cmidrule(lr){3-5} \cmidrule(lr){6-8} \cmidrule(lr){9-11} \cmidrule(lr){12-14}
    & $m$ & 30 & 40 & 50 & 30 & 40 & 50 & 30 & 40 & 50 & 30 & 40 & 50 \\
    \midrule
    \textbf{\dpvoterag{}} & & & & & & & & & & & & & \\
    $\epsilon_\mathrm{token}=1$ & & 
    $0.24$&$0.28$&$\bm{0.29}$& $0.34$&$0.41$&$\bm{0.42}$& $0.34$&$0.41$&$0.42$& $0.34$&$0.41$&$0.42$ \\
    $\epsilon_\mathrm{token}=2$ & & 
    $0.18$&$0.18$&$0.19$& $0.30$&$0.31$&$0.32$& $0.43$&$\bm{0.45}$&$0.44$& $0.43$&$\bm{0.45}$&$0.44$ \\
    $\epsilon_\mathrm{token}=5$ & &
    $0.06$&$0.05$&$0.07$& $0.18$&$0.18$&$0.19$& $0.29$&$0.30$&$0.30$& $0.38$&$0.39$&$0.37$ \\
    \midrule
    \textbf{\dpsparsevoterag{}} & & & & & & & & & & & & & \\
    $\epsilon_\mathrm{token}=1$ & & 
    $0.25$&$0.32$&$0.38$& $0.25$&$0.32$&$0.39$& $0.25$&$0.32$&$0.39$& $0.25$&$0.32$&$0.39$ \\
    $\epsilon_\mathrm{token}=2$ & & 
    $0.37$&$\bm{0.40}$&$\bm{0.40}$& $0.42$&$0.47$&$\bm{0.48}$& $0.43$&$0.53$&$\bm{0.54}$& $0.43$&$0.53$&$0.54$ \\
    $\epsilon_\mathrm{token}=5$ & & 
    $0.25$&$0.25$&$0.26$& $0.43$&$0.44$&$0.41$& $0.51$&$0.53$&$0.51$& $\bm{0.57}$&$\bm{0.57}$&$\bm{0.57}$ \\
    \bottomrule
    \end{tabular}
\end{table*}

\begin{table*}[htbp]
    \caption{Average match accuracy comparison of our algorithms on NQ dataset with Pythia (1.4B) under varying values of total privacy budget $\epsilon_\mathrm{total}$ with different hyperparameters, the number of voters $m$ and $\epsilon_\mathrm{token}$. Bold font values represent the best performance of each algorithm under fixed $\epsilon_\mathrm{total}$.
    We report the means of average match accuracy over three runs.}
    \label{tab:detailed-result-nq-pythia}
    \centering
    \begin{tabular}{l c *{4}{c c c}}
    \toprule
    Algorithm & 
    & \multicolumn{3}{c}{$\epsilon_\mathrm{total} = 5$} & \multicolumn{3}{c}{$\epsilon_\mathrm{total} = 10$} & 
    \multicolumn{3}{c}{$\epsilon_\mathrm{total} = 20$} & \multicolumn{3}{c}{$\epsilon_\mathrm{total} = 40$} \\
    \cmidrule(lr){3-5} \cmidrule(lr){6-8} \cmidrule(lr){9-11} \cmidrule(lr){12-14}
    & $m$ & 30 & 40 & 50 & 30 & 40 & 50 & 30 & 40 & 50 & 30 & 40 & 50 \\
    \midrule
    \textbf{\dpvoterag{}} & & & & & & & & & & & & & \\
    $\epsilon_\mathrm{token}=1$ & & 
    $0.16$&$\bm{0.23}$&$\bm{0.23}$& $0.21$&$0.30$&$\bm{0.32}$& $0.21$&$0.30$&$0.32$& $0.21$&$0.30$&$0.32$ \\
    $\epsilon_\mathrm{token}=2$ & & 
    $0.09$&$0.08$&$0.09$& $0.23$&$0.26$&$0.26$& $0.32$&$\bm{0.35}$&$\bm{0.35}$& $0.32$&$\bm{0.35}$&$\bm{0.35}$ \\
    $\epsilon_\mathrm{token}=5$ & &
    $0.05$&$0.04$&$0.04$& $0.10$&$0.09$&$0.09$& $0.23$&$0.24$&$0.23$& $0.31$&$0.32$&$0.30$ \\
    \midrule
    \textbf{\dpsparsevoterag{}} & & & & & & & & & & & & & \\
    $\epsilon_\mathrm{token}=1$ & & 
    $0.09$&$0.17$&$0.24$& $0.09$&$0.17$&$0.24$& $0.09$&$0.17$&$0.24$& $0.09$&$0.17$&$0.24$ \\
    $\epsilon_\mathrm{token}=2$ & & 
    $0.20$&$0.25$&$\bm{0.26}$& $0.25$&$0.36$&$\bm{0.41}$& $0.26$&$0.37$&$\bm{0.43}$& $0.26$&$0.37$&$0.43$ \\
    $\epsilon_\mathrm{token}=5$ & & 
    $0.14$&$0.14$&$0.15$& $0.27$&$0.28$&$0.27$& $0.39$&$0.41$&$\bm{0.43}$& $0.46$&$\bm{0.51}$&$0.49$ \\
    \bottomrule
    \end{tabular}
\end{table*}

\begin{table*}[htbp]
    \caption{Average match accuracy comparison of our algorithms on Trivia dataset with Llama 3.1 (8B) under varying values of total privacy budget $\epsilon_\mathrm{total}$ with different hyperparameters, the number of voters $m$ and $\epsilon_\mathrm{token}$. Bold font values represent the best performance of each algorithm under fixed $\epsilon_\mathrm{total}$.
    We report the means of average match accuracy over three runs.}
    \label{tab:detailed-result-trivia-llama31}
    \centering
    \begin{tabular}{l c *{4}{c c c}}
    \toprule
    Algorithm & 
    & \multicolumn{3}{c}{$\epsilon_\mathrm{total} = 5$} & \multicolumn{3}{c}{$\epsilon_\mathrm{total} = 10$} & 
    \multicolumn{3}{c}{$\epsilon_\mathrm{total} = 20$} & \multicolumn{3}{c}{$\epsilon_\mathrm{total} = 40$} \\
    \cmidrule(lr){3-5} \cmidrule(lr){6-8} \cmidrule(lr){9-11} \cmidrule(lr){12-14}
    & $m$ & 30 & 40 & 50 & 30 & 40 & 50 & 30 & 40 & 50 & 30 & 40 & 50 \\
    \midrule
    \textbf{\dpvoterag{}} & & & & & & & & & & & & & \\
    $\epsilon_\mathrm{token}=1$ & & 
    $0.75$&$0.77$&$\bm{0.79}$& $0.81$&$0.84$&$\bm{0.85}$& $0.81$&$0.84$&$\bm{0.85}$& $0.81$&$0.84$&$\bm{0.85}$ \\
    $\epsilon_\mathrm{token}=2$ & & 
    $0.51$&$0.51$&$0.51$& $0.78$&$0.78$&$0.78$& $\bm{0.85}$&$\bm{0.85}$&$\bm{0.85}$& $\bm{0.85}$&$\bm{0.85}$&$\bm{0.85}$ \\
    $\epsilon_\mathrm{token}=5$ & &
    $0.20$&$0.21$&$0.21$& $0.51$&$0.51$&$0.51$& $0.73$&$0.74$&$0.74$& $0.83$&$0.82$&$0.82$ \\
    \midrule
    \textbf{\dpsparsevoterag{}} & & & & & & & & & & & & & \\
    $\epsilon_\mathrm{token}=1$ & & 
    $0.72$&$0.83$&$0.88$& $0.72$&$0.83$&$0.88$& $0.72$&$0.83$&$0.88$& $0.72$&$0.83$&$0.88$ \\
    $\epsilon_\mathrm{token}=2$ & & 
    $0.89$&$\bm{0.94}$&$\bm{0.94}$& $0.93$&$\bm{0.96}$&$\bm{0.96}$& $0.93$&$0.96$&$0.96$& $0.93$&$0.96$&$0.96$ \\
    $\epsilon_\mathrm{token}=5$ & & 
    $0.76$&$0.78$&$0.76$& $0.93$&$0.93$&$0.93$& $0.95$&$\bm{0.97}$&$\bm{0.97}$& $0.95$&$\bm{0.97}$&$\bm{0.97}$ \\
    \bottomrule
    \end{tabular}
\end{table*}

\begin{table*}[htbp]
    \caption{Average match accuracy comparison of our algorithms on NQ dataset with Llama 3.1 (8B) under varying values of total privacy budget $\epsilon_\mathrm{total}$ with different hyperparameters, the number of voters $m$ and $\epsilon_\mathrm{token}$. Bold font values represent the best performance of each algorithm under fixed $\epsilon_\mathrm{total}$.
    We report the means of average match accuracy over three runs.}
    \label{tab:detailed-result-nq-llama31}
    \centering
    \begin{tabular}{l c *{4}{c c c}}
    \toprule
    Algorithm & 
    & \multicolumn{3}{c}{$\epsilon_\mathrm{total} = 5$} & \multicolumn{3}{c}{$\epsilon_\mathrm{total} = 10$} & 
    \multicolumn{3}{c}{$\epsilon_\mathrm{total} = 20$} & \multicolumn{3}{c}{$\epsilon_\mathrm{total} = 40$} \\
    \cmidrule(lr){3-5} \cmidrule(lr){6-8} \cmidrule(lr){9-11} \cmidrule(lr){12-14}
    & $m$ & 30 & 40 & 50 & 30 & 40 & 50 & 30 & 40 & 50 & 30 & 40 & 50 \\
    \midrule
    \textbf{\dpvoterag{}} & & & & & & & & & & & & & \\
    $\epsilon_\mathrm{token}=1$ & & 
    $0.49$&$\bm{0.54}$&$\bm{0.54}$& $0.57$&$0.63$&$\bm{0.64}$& $0.57$&$0.63$&$0.64$& $0.57$&$0.63$&$0.64$ \\
    $\epsilon_\mathrm{token}=2$ & & 
    $0.30$&$0.29$&$0.30$& $0.55$&$0.56$&$0.55$& $0.64$&$\bm{0.66}$&$0.65$& $0.64$&$\bm{0.66}$&$0.65$ \\
    $\epsilon_\mathrm{token}=5$ & &
    $0.11$&$0.11$&$0.11$& $0.29$&$0.29$&$0.30$& $0.52$&$0.52$&$0.51$& $0.62$&$0.63$&$0.62$ \\
    \midrule
    \textbf{\dpsparsevoterag{}} & & & & & & & & & & & & & \\
    $\epsilon_\mathrm{token}=1$ & & 
    $0.38$&$0.55$&$0.65$& $0.38$&$0.55$&$0.66$& $0.38$&$0.55$&$0.66$& $0.38$&$0.55$&$0.66$ \\
    $\epsilon_\mathrm{token}=2$ & & 
    $0.64$&$\bm{0.72}$&$\bm{0.72}$& $0.71$&$0.78$&$\bm{0.80}$& $0.71$&$0.79$&$\bm{0.83}$& $0.71$&$0.79$&$0.83$ \\
    $\epsilon_\mathrm{token}=5$ & & 
    $0.48$&$0.49$&$0.48$& $0.74$&$0.75$&$0.74$& $0.79$&$0.80$&$0.79$& $0.85$&$\bm{0.87}$&$0.85$ \\
    \bottomrule
    \end{tabular}
\end{table*}

\begin{table*}[htbp]
    \caption{Average match accuracy comparison of our algorithms on Trivia dataset with Llama 3.2 (1B) under varying values of total privacy budget $\epsilon_\mathrm{total}$ with different hyperparameters, the number of voters $m$ and $\epsilon_\mathrm{token}$. Bold font values represent the best performance of each algorithm under fixed $\epsilon_\mathrm{total}$.
    We report the means of average match accuracy over three runs.}
    \label{tab:detailed-result-trivia-llama32}
    \centering
    \begin{tabular}{l c *{4}{c c c}}
    \toprule
    Algorithm & 
    & \multicolumn{3}{c}{$\epsilon_\mathrm{total} = 5$} & \multicolumn{3}{c}{$\epsilon_\mathrm{total} = 10$} & 
    \multicolumn{3}{c}{$\epsilon_\mathrm{total} = 20$} & \multicolumn{3}{c}{$\epsilon_\mathrm{total} = 40$} \\
    \cmidrule(lr){3-5} \cmidrule(lr){6-8} \cmidrule(lr){9-11} \cmidrule(lr){12-14}
    & $m$ & 30 & 40 & 50 & 30 & 40 & 50 & 30 & 40 & 50 & 30 & 40 & 50 \\
    \midrule
    \textbf{\dpvoterag{}} & & & & & & & & & & & & & \\
    $\epsilon_\mathrm{token}=1$ & & 
    $0.46$&$0.49$&$\bm{0.52}$& $0.54$&$0.57$&$\bm{0.60}$& $0.54$&$0.57$&$0.60$& $0.54$&$0.57$&$0.60$ \\
    $\epsilon_\mathrm{token}=2$ & & 
    $0.27$&$0.27$&$0.28$& $0.51$&$0.52$&$0.52$& $0.60$&$\bm{0.61}$&$0.60$& $0.60$&$\bm{0.61}$&$0.60$ \\
    $\epsilon_\mathrm{token}=5$ & &
    $0.10$&$0.10$&$0.10$& $0.27$&$0.27$&$0.28$& $0.47$&$0.47$&$0.47$& $0.56$&$0.57$&$0.56$ \\
    \midrule
    \textbf{\dpsparsevoterag{}} & & & & & & & & & & & & & \\
    $\epsilon_\mathrm{token}=1$ & & 
    $0.38$&$0.52$&$0.63$& $0.38$&$0.52$&$0.63$& $0.38$&$0.52$&$0.63$& $0.38$&$0.52$&$0.63$ \\
    $\epsilon_\mathrm{token}=2$ & & 
    $0.61$&$0.62$&$\bm{0.64}$& $0.70$&$0.76$&$\bm{0.77}$& $0.71$&$0.78$&$\bm{0.79}$& $0.71$&$0.78$&$0.79$ \\
    $\epsilon_\mathrm{token}=5$ & & 
    $0.43$&$0.43$&$0.45$& $0.66$&$0.64$&$0.66$& $0.76$&$0.75$&$0.76$& $\bm{0.81}$&$0.80$&$0.79$ \\
    \bottomrule
    \end{tabular}
\end{table*}

\begin{table*}[htbp]
    \caption{Average match accuracy comparison of our algorithms on NQ dataset with Llama 3.2 (1B) under varying values of total privacy budget $\epsilon_\mathrm{total}$ with different hyperparameters, the number of voters $m$ and $\epsilon_\mathrm{token}$. Bold font values represent the best performance of each algorithm under fixed $\epsilon_\mathrm{total}$.
    We report the means of average match accuracy over three runs.}
    \label{tab:detailed-result-nq-llama32}
    \centering
    \begin{tabular}{l c *{4}{c c c}}
    \toprule
    Algorithm & 
    & \multicolumn{3}{c}{$\epsilon_\mathrm{total} = 5$} & \multicolumn{3}{c}{$\epsilon_\mathrm{total} = 10$} & 
    \multicolumn{3}{c}{$\epsilon_\mathrm{total} = 20$} & \multicolumn{3}{c}{$\epsilon_\mathrm{total} = 40$} \\
    \cmidrule(lr){3-5} \cmidrule(lr){6-8} \cmidrule(lr){9-11} \cmidrule(lr){12-14}
    & $m$ & 30 & 40 & 50 & 30 & 40 & 50 & 30 & 40 & 50 & 30 & 40 & 50 \\
    \midrule
    \textbf{\dpvoterag{}} & & & & & & & & & & & & & \\
    $\epsilon_\mathrm{token}=1$ & & 
    $0.24$&$\bm{0.27}$&$\bm{0.27}$& $0.31$&$0.38$&$\bm{0.39}$& $0.31$&$0.38$&$0.39$& $0.31$&$0.38$&$0.39$ \\
    $\epsilon_\mathrm{token}=2$ & & 
    $0.15$&$0.15$&$0.12$& $0.30$&$0.31$&$0.27$& $0.42$&$\bm{0.44}$&$0.42$& $0.42$&$\bm{0.44}$&$0.42$ \\
    $\epsilon_\mathrm{token}=5$ & &
    $0.05$&$0.05$&$0.05$& $0.15$&$0.15$&$0.13$& $0.30$&$0.30$&$0.27$& $0.38$&$0.39$&$0.35$ \\
    \midrule
    \textbf{\dpsparsevoterag{}} & & & & & & & & & & & & & \\
    $\epsilon_\mathrm{token}=1$ & & 
    $0.14$&$0.25$&$0.32$& $0.14$&$0.25$&$0.34$& $0.14$&$0.25$&$0.34$& $0.14$&$0.25$&$0.34$ \\
    $\epsilon_\mathrm{token}=2$ & & 
    $0.31$&$\bm{0.33}$&$\bm{0.33}$& $0.41$&$0.49$&$\bm{0.52}$& $0.42$&$0.50$&$\bm{0.53}$& $0.42$&$0.50$&$0.53$ \\
    $\epsilon_\mathrm{token}=5$ & & 
    $0.15$&$0.14$&$0.13$& $0.40$&$0.36$&$0.32$& $0.53$&$0.51$&$\bm{0.53}$& $0.57$&$0.56$&$\bm{0.58}$ \\
    \bottomrule
    \end{tabular}
\end{table*}

\begin{table*}[htbp]
    \caption{Average match accuracy comparison of our algorithms on Trivia dataset with GPT2-XL under varying values of total privacy budget $\epsilon_\mathrm{total}$ with different hyperparameters, the number of voters $m$ and $\epsilon_\mathrm{token}$. Bold font values represent the best performance of each algorithm under fixed $\epsilon_\mathrm{total}$.
    We report the means of average match accuracy over three runs.}
    \label{tab:detailed-result-trivia-gpt2}
    \centering
    \begin{tabular}{l c *{4}{c c c}}
    \toprule
    Algorithm & 
    & \multicolumn{3}{c}{$\epsilon_\mathrm{total} = 5$} & \multicolumn{3}{c}{$\epsilon_\mathrm{total} = 10$} & 
    \multicolumn{3}{c}{$\epsilon_\mathrm{total} = 20$} & \multicolumn{3}{c}{$\epsilon_\mathrm{total} = 40$} \\
    \cmidrule(lr){3-5} \cmidrule(lr){6-8} \cmidrule(lr){9-11} \cmidrule(lr){12-14}
    & $m$ & 30 & 40 & 50 & 30 & 40 & 50 & 30 & 40 & 50 & 30 & 40 & 50 \\
    \midrule
    \textbf{\dpvoterag{}} & & & & & & & & & & & & & \\
    $\epsilon_\mathrm{token}=1$ & & 
    $0.31$&$0.32$&$\bm{0.34}$& $0.36$&$0.38$&$\bm{0.42}$& $0.36$&$0.38$&$0.42$& $0.36$&$0.38$&$0.42$ \\
    $\epsilon_\mathrm{token}=2$ & & 
    $0.22$&$0.23$&$0.23$& $0.37$&$0.38$&$0.37$& $0.44$&$\bm{0.46}$&$\bm{0.46}$& $0.44$&$\bm{0.46}$&$\bm{0.46}$ \\
    $\epsilon_\mathrm{token}=5$ & &
    $0.08$&$0.08$&$0.08$& $0.23$&$0.23$&$0.23$& $0.34$&$0.34$&$0.33$& $0.44$&$0.43$&$0.42$ \\
    \midrule
    \textbf{\dpsparsevoterag{}} & & & & & & & & & & & & & \\
    $\epsilon_\mathrm{token}=1$ & & 
    $0.25$&$0.30$&$0.38$& $0.25$&$0.30$&$0.38$& $0.25$&$0.30$&$0.38$& $0.25$&$0.30$&$0.38$ \\
    $\epsilon_\mathrm{token}=2$ & & 
    $0.37$&$\bm{0.39}$&$0.38$& $0.41$&$0.46$&$\bm{0.47}$& $0.42$&$0.47$&$0.48$& $0.42$&$0.47$&$0.48$ \\
    $\epsilon_\mathrm{token}=5$ & & 
    $0.28$&$0.27$&$0.27$& $0.43$&$0.43$&$0.43$& $0.52$&$\bm{0.53}$&$\bm{0.53}$& $\bm{0.59}$&$0.57$&$0.56$ \\
    \bottomrule
    \end{tabular}
\end{table*}

\begin{table*}[htbp]
    \caption{Average match accuracy comparison of our algorithms on NQ dataset with GPT2-XL under varying values of total privacy budget $\epsilon_\mathrm{total}$ with different hyperparameters, the number of voters $m$ and $\epsilon_\mathrm{token}$. Bold font values represent the best performance of each algorithm under fixed $\epsilon_\mathrm{total}$.
    We report the means of average match accuracy over three runs.}
    \label{tab:detailed-result-nq-gpt2}
    \centering
    \begin{tabular}{l c *{4}{c c c}}
    \toprule
    Algorithm & 
    & \multicolumn{3}{c}{$\epsilon_\mathrm{total} = 5$} & \multicolumn{3}{c}{$\epsilon_\mathrm{total} = 10$} & 
    \multicolumn{3}{c}{$\epsilon_\mathrm{total} = 20$} & \multicolumn{3}{c}{$\epsilon_\mathrm{total} = 40$} \\
    \cmidrule(lr){3-5} \cmidrule(lr){6-8} \cmidrule(lr){9-11} \cmidrule(lr){12-14}
    & $m$ & 30 & 40 & 50 & 30 & 40 & 50 & 30 & 40 & 50 & 30 & 40 & 50 \\
    \midrule
    \textbf{\dpvoterag{}} & & & & & & & & & & & & & \\
    $\epsilon_\mathrm{token}=1$ & & 
    $0.21$&$0.24$&$\bm{0.27}$& $0.25$&$0.28$&$\bm{0.33}$& $0.25$&$0.28$&$0.33$& $0.25$&$0.28$&$0.33$ \\
    $\epsilon_\mathrm{token}=2$ & & 
    $0.18$&$0.19$&$0.19$& $0.30$&$0.29$&$0.29$& $0.35$&$0.35$&$\bm{0.36}$& $0.35$&$0.35$&$\bm{0.36}$ \\
    $\epsilon_\mathrm{token}=5$ & &
    $0.08$&$0.08$&$0.07$& $0.19$&$0.19$&$0.19$& $0.30$&$0.29$&$0.29$& $\bm{0.36}$&$0.34$&$0.34$ \\
    \midrule
    \textbf{\dpsparsevoterag{}} & & & & & & & & & & & & & \\
    $\epsilon_\mathrm{token}=1$ & & 
    $0.14$&$0.19$&$0.24$& $0.14$&$0.19$&$0.24$& $0.14$&$0.19$&$0.24$& $0.14$&$0.19$&$0.24$ \\
    $\epsilon_\mathrm{token}=2$ & & 
    $0.24$&$0.28$&$\bm{0.31}$& $0.28$&$0.33$&$\bm{0.36}$& $0.29$&$0.35$&$0.40$& $0.29$&$0.35$&$0.40$ \\
    $\epsilon_\mathrm{token}=5$ & & 
    $0.18$&$0.18$&$0.16$& $0.34$&$0.33$&$0.35$& $\bm{0.41}$&$0.39$&$\bm{0.41}$& $0.42$&$0.41$&$\bm{0.43}$ \\
    \bottomrule
    \end{tabular}
\end{table*}


\begin{table*}[htbp]
    \caption{Average match accuracy comparison of our algorithms on ChatDoctor dataset with OPT (1.3B) under varying values of total privacy budget $\epsilon_\mathrm{total}$ with different hyperparameters, the number of voters $m$ and $\epsilon_\mathrm{token}$. Bold font values represent the best performance of each algorithm under fixed $\epsilon_\mathrm{total}$.
    We report the means of average match accuracy over three runs.}
    \label{tab:detailed-result-nq-gpt2}
    \centering
    \begin{tabular}{l c *{4}{c c c}}
    \toprule
    Algorithm & 
    & \multicolumn{3}{c}{$\epsilon_\mathrm{total} = 5$} & \multicolumn{3}{c}{$\epsilon_\mathrm{total} = 10$} & 
    \multicolumn{3}{c}{$\epsilon_\mathrm{total} = 20$} & \multicolumn{3}{c}{$\epsilon_\mathrm{total} = 40$} \\
    \cmidrule(lr){3-5} \cmidrule(lr){6-8} \cmidrule(lr){9-11} \cmidrule(lr){12-14}
    & $m$ & 30 & 40 & 50 & 30 & 40 & 50 & 30 & 40 & 50 & 30 & 40 & 50 \\
    \midrule
    \textbf{\dpvoterag{}} & & & & & & & & & & & & & \\
    $\epsilon_\mathrm{token}=1$ & & $0.82$ & $\mathbf{0.83}$ & $\mathbf{0.83}$ & $0.82$ & $0.83$ & $0.83$ & $0.82$ & $\mathbf{0.83}$ & $\mathbf{0.83}$  & $0.82$ & $\mathbf{0.83}$ & $\mathbf{0.83}$  \\
    $\epsilon_\mathrm{token}=2$ & & $0.82$ & $0.82$ & $0.81$ & $0.83$ & $0.83$ & $0.83$ & $\mathbf{0.83}$ & $\mathbf{0.83}$  & $\mathbf{0.83}$ & $\mathbf{0.83}$  & $\mathbf{0.83}$ & $\mathbf{0.83}$  \\
    $\epsilon_\mathrm{token}=5$ & & $0.82$ & $0.82$ & $0.82$ & $\mathbf{0.84}$ & $\mathbf{0.84}$ & $\mathbf{0.84}$ & $0.82$ & $0.82$ & $0.82$ & $\mathbf{0.83}$  & $\mathbf{0.83}$ & $\mathbf{0.83}$ \\
\midrule
    \textbf{\dpsparsevoterag{}} & & & & & & & & & & & & & \\
    $\epsilon_\mathrm{token}=1$ & & $0.79$ & $0.80$ & $0.81$ & $0.79$ & $0.80$ & $0.81$ & $0.79$ & $0.80$ & $0.81$ & $0.79$ & $0.80$ & $0.81$ \\
    $\epsilon_\mathrm{token}=2$ & & $0.83$ & $\mathbf{0.84}$ & $\mathbf{0.84}$ & $0.82$ & $\mathbf{0.84}$ & $\mathbf{0.84}$ & $0.82$ & $0.84$ & $\mathbf{0.85}$ & $0.82$ & $0.84$ & $\mathbf{0.85}$  \\
    $\epsilon_\mathrm{token}=5$ & & $0.13$ & $0.13$ & $0.13$ & $\mathbf{0.84}$ & $\mathbf{0.84}$ & $\mathbf{0.84}$ & $0.84$ &  $0.84$ & $0.84$ &  $\mathbf{0.85}$ & $\mathbf{0.85}$ & $\mathbf{0.85}$ \\
\bottomrule
    \end{tabular}
\end{table*}

\begin{table*}[htbp]
    \caption{Average match accuracy comparison of our algorithms on ChatDoctor dataset with Pythia (1.4B) under varying values of total privacy budget $\epsilon_\mathrm{total}$ with different hyperparameters, the number of voters $m$ and $\epsilon_\mathrm{token}$. Bold font values represent the best performance of each algorithm under fixed $\epsilon_\mathrm{total}$.
    We report the means of average match accuracy over three runs.}
    \label{tab:detailed-result-nq-gpt2}
    \centering
    \begin{tabular}{l c *{4}{c c c}}
    \toprule
    Algorithm & 
    & \multicolumn{3}{c}{$\epsilon_\mathrm{total} = 5$} & \multicolumn{3}{c}{$\epsilon_\mathrm{total} = 10$} & 
    \multicolumn{3}{c}{$\epsilon_\mathrm{total} = 20$} & \multicolumn{3}{c}{$\epsilon_\mathrm{total} = 40$} \\
    \cmidrule(lr){3-5} \cmidrule(lr){6-8} \cmidrule(lr){9-11} \cmidrule(lr){12-14}
    & $m$ & 30 & 40 & 50 & 30 & 40 & 50 & 30 & 40 & 50 & 30 & 40 & 50 \\
    \midrule
    \textbf{DPVoteRAG} & & & & & & & & & & & & & \\
    $\epsilon_\mathrm{token}=1$ & & $0.82$ & $0.84$ & $\bm{0.85}$ & $0.82$ & $0.84$ & $\bm{0.85}$ & $0.82$ & $0.84$ & $\bm{0.85}$ & $0.82$ & $0.84$ & $\bm{0.85}$ \\
    $\epsilon_\mathrm{token}=2$ & & $0.82$ & $0.82$ & $0.82$ & $0.84$ & $\bm{0.85}$ & $\bm{0.85}$ & $0.84$ & $\bm{0.85}$ & $\bm{0.85}$ & $0.84$ & $\bm{0.85}$ & $\bm{0.85}$ \\
    $\epsilon_\mathrm{token}=5$ & & $0.84$ & $0.84$ & $0.84$ & $0.81$ & $0.82$ & $0.81$ & $0.84$ & $0.84$ & $0.84$ & $\bm{0.85}$ & $\bm{0.85}$ & $\bm{0.85}$ \\
\midrule
    \textbf{DPSparseVoteRAG} & & & & & & & & & & & & & \\
    $\epsilon_\mathrm{token}=1$ & & $0.79$ & $0.80$ & $0.81$ & $0.79$ & $0.80$ & $0.81$ & $0.79$ & $0.80$ & $0.81$ & $0.79$ & $0.80$ & $0.81$ \\
    $\epsilon_\mathrm{token}=2$ & & $0.82$ & $0.82$ & $\bm{0.83}$ & $0.83$ & $0.84$ & $\bm{0.85}$ & $0.83$ & $\bm{0.85}$ & $\bm{0.85}$ & $0.83$ & $\bm{0.85}$ & $\bm{0.85}$ \\
    $\epsilon_\mathrm{token}=5$ & & $0.82$ & $0.82$ & $0.82$ & $0.83$ & $0.83$ & $0.83$ & $\bm{0.85}$ & $\bm{0.85}$ & $\bm{0.85}$ & $\bm{0.85}$ & $\bm{0.85}$ & $\bm{0.85}$ \\    
\bottomrule
    \end{tabular}
\end{table*}

\begin{table*}[htbp]
    \caption{Average match accuracy comparison of our algorithms on ChatDoctor dataset with Llama 3.1 (8B) under varying values of total privacy budget $\epsilon_\mathrm{total}$ with different hyperparameters, the number of voters $m$ and $\epsilon_\mathrm{token}$. Bold font values represent the best performance of each algorithm under fixed $\epsilon_\mathrm{total}$.
    We report the means of average match accuracy over three runs.}
    \label{tab:detailed-result-nq-gpt2}
    \centering
    \begin{tabular}{l c *{4}{c c c}}
    \toprule
    Algorithm & 
    & \multicolumn{3}{c}{$\epsilon_\mathrm{total} = 5$} & \multicolumn{3}{c}{$\epsilon_\mathrm{total} = 10$} & 
    \multicolumn{3}{c}{$\epsilon_\mathrm{total} = 20$} & \multicolumn{3}{c}{$\epsilon_\mathrm{total} = 40$} \\
    \cmidrule(lr){3-5} \cmidrule(lr){6-8} \cmidrule(lr){9-11} \cmidrule(lr){12-14}
    & $m$ & 30 & 40 & 50 & 30 & 40 & 50 & 30 & 40 & 50 & 30 & 40 & 50 \\
    \midrule
    \textbf{\dpvoterag{}} & & & & & & & & & & & & & \\
    $\epsilon_\mathrm{token}=1$ & & $0.83$ & $\mathbf{0.84}$ & $\mathbf{0.84}$ & $0.83$ & $\mathbf{0.84}$ & $\mathbf{0.84}$ & $0.83$ & $\mathbf{0.84}$ & $\mathbf{0.84}$ & $0.83$ & $0.84$ & $0.84$ \\
    $\epsilon_\mathrm{token}=2$ & & $0.83$ & $0.83$ & $0.83$ & $\mathbf{0.84}$ & $\mathbf{0.84}$& $\mathbf{0.84}$  & $\mathbf{0.84}$ & $\mathbf{0.84}$& $\mathbf{0.84}$ & $0.84$ & $0.84$ & $0.84$ \\
    $\epsilon_\mathrm{token}=5$ & & $0.82$ & $0.82$ & $0.82$ & $\mathbf{0.84}$ & $\mathbf{0.84}$& $\mathbf{0.84}$  &$\mathbf{0.84}$ & $\mathbf{0.84}$& $\mathbf{0.84}$ & $\mathbf{0.85}$ & $0.84$ & $0.84$ \\
\midrule
    \textbf{\dpsparsevoterag{}} & & & & & & & & & & & & & \\
    $\epsilon_\mathrm{token}=1$ & & $0.82$ & $0.83$ & $0.82$ & $0.82$ & $0.83$ & $0.82$ & $0.82$ & $0.83$ & $0.82$ & $0.82$ & $0.83$ & $0.82$ \\
    $\epsilon_\mathrm{token}=2$ & & $\mathbf{0.84}$ & $\mathbf{0.84}$ & $\mathbf{0.84}$ & $\mathbf{0.85}$ & $\mathbf{0.85}$ & $\mathbf{0.85}$ & $0.85$ & $0.85$ & $0.85$ & $0.85$ & $0.85$ & $0.85$ \\
    $\epsilon_\mathrm{token}=5$ & & $0.46$ & $0.45$ & $0.35$ & $0.84$ & $\mathbf{0.85}$ & $0.84$ & $0.85$ & $\mathbf{0.86}$ & $0.85$ & $\mathbf{0.86}$ & $\mathbf{0.86}$ & $\mathbf{0.86}$ \\    
    \bottomrule
    \end{tabular}
\end{table*}
\newpage